\documentclass[twocolumn,trackchanges]{aastex631}

\makeatletter
\let\frontmatter@title@above\relax
\makeatother

\usepackage{amsmath,amssymb,amstext}

\usepackage[all]{hypcap} 
\usepackage[T1]{fontenc} 
\usepackage[caption=false]{subfig}
\usepackage{booktabs}

\usepackage{xcolor}
\usepackage{graphicx}

\usepackage{amsthm}
\usepackage{hyperref}

\newcommand{\subsubsubsection}[1]{%
  \vspace{1.0\baselineskip}%
  \noindent\textit{#1}\par\nobreak%
  \vspace{0.5\baselineskip}%
}

\newcommand{\M}{{\cal M}}

\shortauthors{Giron, Krief, Stone \& Steinberg}
\numberwithin{equation}{section}

\graphicspath{{./}}

\begin{document}

\title{Multigroup Radiation Diffusion on a Moving Mesh: Implementation in \textsc{rich} and Application to Tidal Disruption Events}
\shorttitle{Multigroup Radiation Diffusion on a Moving Mesh}

\author[0000-0003-2234-1319]{Itamar Giron}
\affiliation{Racah Institute of Physics, The Hebrew University, 9190401 Jerusalem, Israel}
\email{itamar.giron@mail.huji.ac.il}

\author[0000-0002-7093-6467]{Menahem Krief}
\email{menahem.krief@mail.huji.ac.il}
\affiliation{Racah Institute of Physics, The Hebrew University, 9190401 Jerusalem, Israel}

\author[0000-0002-4337-9458]{Nicholas C. Stone}
\email{nicholas.stone@mail.huji.ac.il}
\affiliation{Racah Institute of Physics, The Hebrew University, 9190401 Jerusalem, Israel}
\affiliation{Department of Astronomy, University of Wisconsin, Madison, WI 53706, USA}

\author[0000-0003-0053-0696]{Elad Steinberg}
\email{elad.steinberg@mail.huji.ac.il}
\affiliation{Racah Institute of Physics, The Hebrew University, 9190401 Jerusalem, Israel}

\begin{abstract}
Radiation-hydrodynamics (RHD) determines the bulk evolution and observable emission in a wide variety of high-energy astrophysical phenomena.  Due to their complexity, RHD problems must usually be studied through numerical simulation.  We have extended the publicly available \textsc{rich} code, which previously solved the equations of RHD in the limit of grey flux-limited diffusion (FLD), to operate with a multigroup FLD solver.  \textsc{rich} is a semi-Lagrangian code that solves the equations of RHD on an unstructured moving mesh, and is the first multigroup RHD moving mesh code, making it uniquely applicable to problems with extreme dynamic range and dynamically important radiation forces.  We validate our multigroup module against multiple analytic benchmarks, including a novel test of the RHD Doppler term.  The computational efficiency of the code is aided by a novel scheme to accelerate convergence in optically thick cells by limiting the absorption coefficients.  Finally, we apply multigroup \textsc{rich} in a pilot three dimensional study of a stellar tidal disruption event (TDE), using a $10^4 M_\odot$ intermediate-mass black hole.  Our simulations self-consistently produce a bright early-time X-ray flash prior to peak optical/UV light, in qualitative agreement with post-processing of (grey) \textsc{rich} simulations of supermassive black hole TDEs, as well as X-ray observations of the TDE AT 2022dsb.

\end{abstract}

\section{Introduction}


Radiation-hydrodynamics (RHD) is central to a multitude of astrophysical processes.  Energy transport in stellar interiors \citep{KippenhahnWeigert13}, and the flow of matter in \citep{YuanNarayan14, Jiang+19} and out \citep{Fabian99, Thompson+15} of active galactic nuclei exemplify quasi-stationary problems in RHD, for which it is possible to construct approximate analytic or semi-analytic solutions.  Conversely, other important astrophysical phenomena occur in the {\it dynamical} limit of RHD, such as the cooling and fragmentation of gas in star-forming regions \citep{Krumholz+07a, Grudic+21}, or the evolution of complex transients around black holes \citep{Bonnerot+21, steinberg_streamdisk_2024, Brutman+24, Huang+25}.  For dynamical problems such as these, it is uncommon for analytic methods to be practical, and the natural theoretical approach is to discretize and simulate the equations of RHD on a computer.


A variety of numerical methods exist to simulate problems in radiation hydrodynamics \citep{Castor_2004}.  Standard hydrodynamics solvers can be coupled to various discretizations of the equation of radiation transfer.  The most common approach, flux-limited diffusion (FLD), evolves a one-moment closure of the transfer equation, assuming that radiation transfer is diffusive in optically thick regimes, while using a flux limiter to approximate free streaming motion in optically thin regions \citep{Pomeraning-Levermore}.  FLD is generally an efficient approach to RHD; other algorithms, such as the M1 two-moment closure \citep{Levermore84, AubertTeyssier08, Sadowski+13}, the variable Eddington tensor method \citep{Stone+92, HayesNorman03, Jiang+12}, Monte Carlo schemes (\cite{fleck_implicit_1971,gentile2001implicit,wollaber2016four,noebauer2019monte,steinberg2022new,steinberg2023frequency}) or $S_{N}$ and $P_{N}$ deterministic schemes (\cite{Brunner2002Froms,mcclarren2010theoretical,till2018application}), will usually attain greater physical accuracy at the expense of greater computational cost.  Aside from the choice of discretization, a second axis of differentiation between RHD solvers is the treatment of photon spectra.  Some algorithms work in the grey, or single-frequency, approximation, while other multigroup solvers operate on multiple different bins of photons, which propagate differently due to the frequency-dependent opacity of the matter they interact with.


In this paper, we take the existing \textsc{rich} radiation hydrodynamics code and upgrade its radiation transfer module from grey to multigroup FLD.  \textsc{rich} is a publicly available code developed originally for finite-volume semi-Lagrangian (moving-mesh) hydrodynamics \citep{yalinewich_rich_2015, Steinberg+15, Steinberg+16}.  Moving mesh algorithms for hydrodynamics add computational expense (e.g. repeated tessellation of an unstructured mesh) but can avoid common pitfalls of finite mass schemes while retaining their strengths in problems with enormous dynamic range or highly supersonic flow.  Cold flow accretion into galaxies \citep{Nelson+13}, dynamically cold shearing flows \citep{ZierSpringel22}, and stellar tidal disruption events \citep{steinberg_streamdisk_2024} are examples of problems that pose severe computational challenges for both Eulerian, static mesh schemes and finite mass approaches, but which have been fruitfully tackled using moving-mesh codes.  



Indeed, tidal disruption events (TDEs) are one of the main use cases for \textsc{rich} \citep{Krolik+20}, both in its original, purely hydrodynamic form, and in its newer (grey) RHD version.  \textsc{rich} has already been used to perform notable global simulations of TDEs, including the first end-to-end simulation of an astrophysically realistic disruption \citep{steinberg_streamdisk_2024}, and the most detailed series of global TDE convergence tests to date \citep{Martire+25}.  However, these studies have been limited by the grey radiation transfer module in the existing version of \textsc{rich}, which has prevented self-consistent modeling of emergent spectra.  In past \textsc{rich} simulations of TDEs, spectral properties have either been ignored, or have been computed through simplified post-processing schemes \citep{steinberg_streamdisk_2024}.  The primary goal of this paper is to implement multigroup FLD radiation transfer in \textsc{rich}, enabling more sophisticated simulations of high-energy astrophysical processes, including TDEs, stellar collisions \citep{Brutman+24} and novae \citep{SteinbergMetzger18}.  A secondary goal of this paper is to properly document the \textsc{rich} FLD algorithm, which had not previously been written up in a code paper.  Only a small number of moving-mesh radiation hydrodynamics codes exist, such as AREPO-RT \citep{Kannan+19}, MANGA \citep{Chang+20}, and AREPO-IDORT \citep{Ma+25}, all of which operate in the grey limit of radiation transfer, making this the first multigroup moving-mesh RHD code.


In Section~\ref{sec:radiation_diffusion_equation}, we review the monochromatic, comoving, flux‑limited diffusion (FLD) equation and its coupling to material energy and momentum. 
Section~\ref{sec:multigroup_equation} derives the multigroup FLD equations 
from the monochromatic form. Section~\ref{sec:implementation} details the \textsc{{rich}} implementation, including the spatial discretization and the fully implicit time integration for the coupled material–radiation system, and outlines the solution strategy for the resulting large linear system. Section~\ref{sec:numerical_Results} presents verification tests (Marshak waves, a radiative shock, comparisons with multigroup Monte Carlo, and a Doppler‑term test) and an application to a tidal disruption event (TDE), demonstrating the method’s applicability to astrophysical flows. We summarize in Section~\ref{sec:conclusions} and discuss future work.

\section{The Monochromatic Radiation Diffusion Equation} \label{sec:radiation_diffusion_equation}

We adopt the frequency‑dependent, comoving‑frame flux‑limited diffusion (FLD) equation with elastic scattering (i.e. scattering contributes to the transport opacity but does not exchange energy with the material).  The spectral radiation energy density $E_\nu(\mathbf{x},t)$, which has units of energy per unit photon-energy per unit volume, evolves according to  (\citealt{Castor_2004,mihalas_foundations_1999,winslow1995multifrequency}): 
\begin{equation}\label{eq:multifrequency}
\begin{split}
&\frac{\partial E_\nu}{\partial t} + \nabla\cdot\left(\vec{v}E_\nu\right) = -\frac{1-R_{2,\nu}}{2}E_\nu\nabla\cdot\vec{v} + \nabla\cdot\left(D_\nu\nabla E_\nu\right) \\
& + c\kappa_\nu\left(b_\nu U_m-E_\nu\right)
+\frac{\partial\left(\nu E_\nu\left(1-R_{2,\nu}\right)/2\right)}{\partial\nu}\nabla\cdot\vec{v},
\end{split}
\end{equation}
where $\vec{v}$ is the fluid velocity, $\kappa_\nu$ is the monochromatic absorption opacity, $\nu$  is the radiation frequency (in energy units),  $U_m\equiv aT^4$ is the blackbody radiation energy density at the material temperature, $a$ is the radiation constant, and $T$ is the material temperature.  The normalized Planck function is
\begin{equation}
b_\nu \equiv B_\nu(T)/U_m 
\end{equation}
such that $\int_0^\infty b_\nu\,d\nu = 1$, and
\begin{equation}
B_\nu(T) = \frac{8\pi\nu^3}{h^3c^3}\frac{1}{\exp\left(\frac{\nu}{k_B T}\right)-1}
\end{equation}
is the Planck spectrum in energy-density form.  We note that $B_\nu$ here has units of energy density per photon-energy, and is normalized such that $\intop_0^\infty B_\nu(T)d\nu = U_m$. In the flux–limited diffusion approximation, the effective radiative diffusion coefficient $D_\nu$ is given by 
\begin{equation}
   D_\nu=\frac{c\lambda_\nu}{\kappa_{t,\nu}},
\end{equation}
where $\kappa_{t,\nu}$ is the total opacity (absorption + scattering) of a photon of frequency $\nu$, and  $\lambda_\nu$ is a flux limiter. We employ the Levermore–Pomraning flux-limiter \citep{Pomeraning-Levermore}, given by:
\begin{equation}
\begin{split}
   & 
   \lambda_\nu = \frac{1}{R_\nu}\left(\coth{R_\nu} - \frac{1}{R_\nu}\right) \\
   & R_\nu = \frac{\left|\nabla E_\nu\right|}{\kappa_{t,\nu}E_\nu},\
   R_{2,\nu}=\lambda_\nu+\lambda_\nu^2R_\nu^2
\end{split}
\end{equation}

The terms in Eq. \ref{eq:multifrequency} encode different forms of work and energy transport, and their physical meanings can be understood in the following ways: $\nabla\cdot\left(\vec{v}E_\nu\right)$ describes comoving advection of the radiation field in the fluid flow, $-\frac{1}{2}(1-R_{2,\nu})E_\nu\nabla\cdot\vec{v}$ describes a mechanical $P{\rm d}V$ exchange, $\nabla\cdot\left(D_\nu\nabla E_\nu\right)$ describes diffusion of radiation energy, $c\kappa_\nu\left(b_\nu U_m-E_\nu\right)$ couples the radiation energy density to the material energy density through emission/absorption, and $\frac{\partial}{\partial\nu}\left(\nu E_\nu\left(1-R_{2,\nu}\right)/2\right)\nabla\cdot\vec{v}$ encodes the Doppler shift. 

The total material energy density $E_m$ couples to the radiation field via the terms  (\citealt{Castor_2004,mihalas_foundations_1999}):
\begin{equation}\label{eq:Energy}
\begin{split} 
   & \frac{\partial E_m}{\partial t} +\nabla\cdot\left[\left(E_m+P\right)\vec{v}\right] = \\
   &  -\vec{v}\cdot \intop_0^\infty\lambda_\nu\nabla E_\nu d\nu - c\intop_0^\infty\kappa_\nu(b_\nu U_m-E_\nu)d\nu.
\end{split}
\end{equation}
The radiation force enters the momentum density equation via the term
\begin{equation}\label{eq:momentum}
    \frac{\partial(\rho\vec{v})}{\partial t} +\nabla\cdot\left(\rho\vec{v}\vec{v}\right)= - \nabla P-\intop_0^\infty\lambda_\nu \nabla E_\nu d\nu .
\end{equation}
\textsc{rich} advances Eqs.~\eqref{eq:multifrequency}–\eqref{eq:momentum} with operator splitting, and the purely hydrodynamic terms \(\nabla\cdot[(E_m+P)\vec v]\), \(\nabla\!\cdot(\rho\vec v\vec v)\), and \(\nabla P\) are updated in the hydro step and  are unaffected by the distinction between grey and multigroup FLD. In what follows we focus on the radiation–matter system, omitting the hydro-only terms from the notation.

\section{The Multigroup Diffusion Equation} \label{sec:multigroup_equation}

We partition the spectrum into a grid of $G$ groups with interfaces at
$\nu_{ 1/2} < \cdots < \nu_{G+ 1/2}$. The group centers introduce a degree of freedom, and we define them using the arithmetic mean $\nu_g \equiv (\nu_{g- 1/2}+\nu_{g+ 1/2})/2$.

We adopt the group-integrated radiation variables $E_g$ and $b_g$ 
between interfaces $\nu_{g\pm 1/2}$, centered at $\nu_g$:

\begin{equation}
E_g \;\equiv\; \int_{\nu_{g- 1/2}}^{\nu_{g+ 1/2}} E_\nu\,d\nu,
\qquad
b_g \;\equiv\; \int_{\nu_{g- 1/2}}^{\nu_{g+ 1/2}} b_\nu\,d\nu,
\end{equation}

The group Planck integrals $b_g$ are approximated using a truncated series (see appendix \ref{sec:computing_group_planck_integrals}). We assume that the frequency grid covers the Planckian well for all relevant material temperatures, that is 

$$\sum_{g=1}^G b_g(T) \approx 1$$

Integrating Eq. \ref{eq:multifrequency} over the range $[\nu_{g- 1/2},\nu_{g+ 1/2}]$ yields  the multigroup radiation diffusion equation:
\begin{equation}
    \label{eq:multigroup}
    \begin{split}
    &\frac{\partial E_g}{\partial t} 
    + \nabla \cdot \left( \vec{v}E_g\right)
    = - \frac{1-R_{2,g}}{2}E_g\nabla\cdot\vec{v}
    \\
    & +\nabla\cdot\left(D_g\nabla E_g \right) + c\kappa_{P,g}\left(b_g U_m - E_g\right) \\
    & +\nabla\cdot\vec{v} 
    \intop_{\nu_{g- 1/2}}^{\nu_{g+ 1/2}}\frac{\partial\left(\nu E_\nu(1-R_{2,\nu})/2\right)}{\partial\nu}d\nu,
    \end{split}
\end{equation}
where the diffusion coefficient in group $g$ is:
\begin{equation}
D_g \equiv \frac{c\lambda_g}{\kappa_{R,g}},
\end{equation}
with the group flux limiter:
\begin{equation}
\begin{split}
   &\ \lambda_g = \frac{1}{R_g}\left(\coth{R_g} - \frac{1}{R_g}\right) \\
   &R_g = \frac{\left|\nabla E_g\right|}{\kappa_{R,g}E_g},\ 
   R_{2,g}=\lambda_g+\lambda_g^2R_g^2,
\end{split}
\end{equation}
and the group average Rosseland mean opacity:
\begin{equation}
    \kappa_{R, g}^{-1} = \frac{\intop_{\nu_{g- 1/2}}^{\nu_{g+ 1/2}}\kappa_{t,\nu }^{-1}\left[\partial B_\nu/\partial T\right]d\nu}{\intop_{\nu_{g- 1/2}}^{\nu_{g+ 1/2}}\partial B_\nu/\partial Td\nu}.
\end{equation}
The average group absorption ("Planck") opacity is:
\begin{equation}
    \kappa_{P,g} = \frac{\intop_{\nu_{g- 1/2}}^{\nu_{g+ 1/2}}\kappa_\nu B_\nu(T)d\nu }{\intop_{\nu_{g- 1/2}}^{\nu_{g+ 1/2}}B_\nu(T)d\nu}.
\end{equation}
Splitting the terms in Eq.  \ref{eq:Energy} over the energy groups gives
\begin{equation}\label{eq:energy_mg}
    \frac{\partial E_m}{\partial t} = 
    -\vec{v}\cdot\sum_{g'}\lambda_{g'}\nabla E_{g'} + c\left(\sum_{g'}\kappa_{P, g'}E_{g'}- \kappa_PU_m\right)
\end{equation}
where
\begin{equation}
\kappa_P=\sum_{g'}\kappa_{P, g'}b_{g'},
\end{equation}
is the total Planck opacity.  Here and below we use a primed index \(g'\) for group sums to distinguish the dummy summation index from the specific group \(g\) appearing in the multigroup transport equation, this avoids ambiguity later when substituting the material–radiation coupling back into Eq.~\eqref{eq:multigroup}.
We separate the material energy equation into an internal‑energy update,
\begin{equation}\label{eq:internal_energy_mg}
    \frac{\partial u_{m}}{\partial t}=c\left(\sum_{g'}\kappa_{P, g'}E_{g'}-\kappa_{P}U_{m}\right)
\end{equation}
and a kinetic‑energy exchange:
\begin{equation}
    \label{eq:kinetic_energy_mg}
    \frac{\partial K}{\partial t}=-\vec{v}\cdot\sum_{g'}\lambda_{g'}\nabla E_{g'}.
\end{equation}
Similarly, splitting the momentum coupling (Eq. \ref{eq:momentum}) over the energy groups yields
\begin{equation}
    \label{eq:momentum_mg}
    \frac{\partial (\rho\vec{v})}{\partial t} = - \sum_{g'}\lambda_{g'}\nabla E_{g'} .
\end{equation}
Equations \ref{eq:multigroup}-\ref{eq:momentum_mg} define the coupled multigroup radiation diffusion equations (modulo the hydro-only terms). 

\section{Implementation in \textsc{rich}} \label{sec:implementation}

\textsc{rich} first advances the Euler equations using a second-order Godunov scheme, then moves the Voronoi mesh according to the velocity field from the hydrodynamic step \citep{yalinewich_rich_2015}, and finally solves the radiation–transfer equations, either in the mixed-frame grey diffusion approximation \citep{steinberg_streamdisk_2024} or in the multigroup diffusion approximation.

The multigroup diffusion step is itself split into three substages. 
\begin{enumerate}
    \item Advection: we apply the comoving advection contribution $\nabla\cdot\left(\vec{v}E_g\right)$ using the mass fluxes from the solution of the Riemann problem. This is performed during the hydro step.
    \item Radiation Step: using an implicit finite-volume scheme, we solve the remaining terms of the multigroup system
    : mechanical $P\mathrm{d}V$ work, diffusion, emission/absorption, and Doppler shifts.  Taken together, the remaining terms define the following differential equation for each group:
\begin{equation}
    \label{eq:the_rest_of_mg}
    \begin{split}
    & \frac{\partial E_g}{\partial t} 
    = - \frac{1-R_{2,g}}{2}E_g\nabla\cdot\vec{v}
     \\
    & +\nabla\cdot\left(D_g\nabla E_g \right) + c\kappa_{P,g}\left(b_g U_m - E_g\right) \\
    &+\intop_{\nu_{g- 1/2}}^{\nu_{g+1/2}}\frac{\partial\left(\nu E_\nu (1-R_{2,\nu})/2\right)}{\partial\nu}d\nu\nabla\cdot\vec{v}.
    \end{split}
\end{equation}
This is solved together with the material energy equation (Eq. \ref{eq:internal_energy_mg}).
    \item Momentum/Kinetic energy: we update the momentum and kinetic energy of the gas using the new radiation field.
\end{enumerate}

\subsection{Discretization of the Multigroup Equation}

Integrating  Eq. \ref{eq:the_rest_of_mg}  over the volume $V_i$ of cell $i$ yields:
\begin{equation}
\begin{split}
    & V_i\frac{\partial E_{g,i}}{\partial t}  = 
        - \frac{1-R_{2,g,i}}{2}E_{g,i}\intop_{V_i}\nabla\cdot\vec{v}dV \\
    & +\intop_{V_i}\nabla\cdot\left(D_g\nabla E_g \right)dV 
    + cV_i\kappa_{P,g,i}\left(b_{g,i} U_{m,i} - E_{g,i}\right) \\
    & +\intop_{\nu_{g- 1/2}}^{\nu_{g+ 1/2}}\frac{\partial\left(\nu E_{\nu,i}(1-R_{2,\nu,i})/2\right)}{\partial\nu}d\nu \intop_{V_i}\nabla\cdot\vec{v}dV.
\end{split}
\end{equation}
Here $E_{g,i} = \frac{1}{V_i}\intop_{V_i}E_{g}dV$ is the volume-averaged group radiation energy density. Using the divergence theorem, we can write
\begin{equation}
    \label{eq:int_div_V}
    \intop_{V_i}\nabla\cdot\vec{v}dV = \intop_{\partial V_i} \vec{v}\cdot d\vec{A} = \sum_{j}^\text{neighbors} \vec{A}_{ij}\cdot\vec{v}_{ij},
\end{equation}
where $\partial V_i$ is the boundary of cell $i$, the sum is across all neighbors $j$ of cell $i$, $\vec{v}_{ij}=\frac{\vec{v}_i + \vec{v}_j}{2}$ is the face-averaged velocity, and $\vec{A}_{ij}$ points in the direction of the outward face normal and has a magnitude equal to the area of the face bordering both cell $i$ and cell $j$. 

For the diffusion term,
\begin{equation}
   \intop_{V_{i}}\nabla\cdot\left(D_{g}\nabla E_{g}\right)dV=\sum_{j}^\text{neighbors}\vec{A}_{ij}\cdot\left(D_{g,ij}\nabla E_{g,ij}\right),
\end{equation}
with the face averaged quantities defined as:
\begin{equation}
\begin{split}
& T_{ij} = \left(\frac{T_i^4+T_j^4}{2}\right)^{1/4},\ D_{g,ij}=\frac{c\lambda_{g,ij}}{\kappa_{R,g,ij}} \\ \ & \kappa_{R,g,ij}=\frac{2}{\kappa_{R,g,i}^{-1}+\kappa_{R,g,j}^{-1}},\ R_{g,ij}=\frac{\left|\nabla E_{g,ij}\right|}{\kappa_{R,g,ij}E_{g,ij}} \\
&E_{g,ij} = \frac{E_{g,i} + E_{g,j}}{2}.
\end{split}
\end{equation}
In the calculation of $\kappa_{R,g,ij}$, $\kappa_{R,g,i}$ and $\kappa_{R,g,j}$, the temperature $T_{ij}$ was used in tandem with their respective cell densities \citep{Till_T4}.

The face-centered energy density gradient is calculated using the center-of-mass position $\overrightarrow {CM}_i$:
\begin{equation}
   \nabla E_{g,ij}=\frac{E_{g,j}-E_{g,i}}{\left\Vert \overrightarrow {CM}_{j}-\overrightarrow {CM}_{i}\right\Vert ^{2}}\left(\overrightarrow {CM}_{j}-\overrightarrow {CM}_{i}\right).
\end{equation}

\subsubsection{Doppler Term}

In this subsection we detail our treatment of the Doppler term,
\begin{equation}    
-\nabla\cdot\vec{v}\frac{\partial\left(\nu E_{\nu}(1-R_{2,\nu})/2\right)}{\partial\nu}.
\label{eq:doppler_only}
\end{equation}
Defining the negative divergence of the velocity as $\alpha\equiv -\nabla\cdot \vec{v}$ and integrating Eq. \ref{eq:doppler_only} over group $g$ results in

\begin{equation}
    \begin{split}
    & \alpha\nu_{g+1/2}\left(\frac{1-R_{2,\nu_{g+1/2}}}{2}\right)E_{\nu_{g+1/2}}\\ 
    &-\alpha\nu_{g-1/2}\left(\frac{1-R_{2,\nu_{g-1/2}}}{2}\right)E_{\nu_{g-1/2}}
    \end{split}
\end{equation}

The interface values (in frequency space) $E_{\nu_{g\pm 1/2}}$ are reconstructed with an upwind, slope–limited scheme. Notice that $E_{\nu_{g\pm 1/2}}$ are energy density \textbf{per unit frequency}.  The left slope  $m_{L,g}$ and the right slope $m_{R,g}$ are: 
\begin{equation}
    \begin{split}
        & m_{L,g} = \frac{E_{\nu_{g}}-E_{\nu_{g-1}}}{\Delta\nu_{g-1}}=\frac{{E_{g}}/{\Delta_{g}}-{E_{g-1}}/{\Delta_{g-1}}}{\Delta\nu_{g-1}} \\
        & m_{R,g}=\frac{E_{\nu_{g+1}}-E_{\nu_{g}}}{\Delta\nu_g}=\frac{{E_{g+1}}/{\Delta_{g+1}}-{E_{g}}/{\Delta_{g}}}{\Delta\nu_{g}}.
    \end{split}
\end{equation}
Here $\Delta_{g}=\nu_{g+1/2}-\nu_{g-1/2}$ is the energy group width and $\Delta\nu_{g}=\nu_{g+1}-\nu_{g}$ is the energy difference between the group centers.

The slope-limiter $\phi(r_g)$ is a function of the group slope ratio $r_{g}=\frac{m_{L,g}}{m_{R,g}}$. In our implementation we use the Superbee slope-limiter (\citealt{Superbee}, see section \ref{sec:doppler_test}), but other limiters can be used as well. The Superbee slope-limiter is given by
\begin{equation}
    \label{eq:superbee_slope_limiter}
    \phi(r) = \max \left(0, \min \left(2r, 1\right),\min\left(r,2\right)\right).
\end{equation}
Since compression ($\alpha \geq 0,\ \nabla\cdot v \leq0$) blue-shifts the photons (i.e. advects them to higher $\nu$), and expansion red-shifts the photons (i.e. advects to lower $\nu$), the upwind direction switches sides.  More concretely: if we consider the $\alpha \geq 0,\ \nabla\cdot v \leq0$ case, the energy groups get blue-shifted, thus the upwind direction is to the lower energies, and $E_{\nu_{g\pm 1/2}}$ are reconstructed by interpolation from the left side,
\begin{equation}
    \begin{split}   
        & E_{\nu_{g+1/2}}={E_{g}}/{\Delta_{g}}+\frac{\Delta_{g}\phi\left(r_{g}\right)}{2\Delta\nu_{g}}\left(\frac{E_{g+1}}{\Delta_{g+1}}-\frac{E_{g}}{\Delta_{g}}\right) \\
        & E_{\nu_{g-1/2}}=\frac{E_{g-1}}{\Delta_{g-1}}+\frac{\Delta_{g-1}\phi\left(r_{g-1}\right)}{2\Delta\nu_{g-1}}\left(\frac{E_{g}}{\Delta_{g}}-\frac{E_{g-1}}{\Delta_{g-1}}\right).
    \end{split}
\end{equation}
Inserting the above into the Doppler term,
\begin{equation}\label{eq:doppler_compression_1}
\begin{split}
   & \alpha\nu_{g+1/2}\left(\frac{1-R_{2,\nu_{g+1/2}}}{2}\right)\Bigg[\frac{E_{g}}{\Delta_{g}} \\
   &\qquad\qquad\quad\quad\quad\quad+\frac{\Delta_{g}} {2}\phi\left(r_{g}\right)\frac{{E_{g+1}}/{\Delta_{g+1}}-{E_{g}}/{\Delta_{g}}}{\Delta\nu_{g}}\Bigg] \\
    &-\alpha\nu_{g-1/2}\left(\frac{1-R_{2,\nu_{g-1/2}}}{2}\right)\Bigg[\frac{E_{g-1}}{\Delta_{g-1}}\\
    &\qquad\qquad\quad\quad\ +\frac{\Delta_{g-1}\phi\left(r_{g-1}\right)}{2}\frac{{E_{g}}/{\Delta_{g}}-{E_{g-1}}/{\Delta_{g-1}}}{\Delta\nu_{g-1}}\Bigg].
\end{split}
\end{equation}
Here we took $R_{2,\nu_{g+1/2}} = R_{2,g}$. Notice that when $\phi= 0$, Eq. \ref{eq:doppler_compression_1} becomes the simple upwind method. 

Integrating over the volume of a cell simply adds an $i$ index to the energy groups, the coefficients, and makes
$$ 
\alpha_i = - \sum_{j}^\text{neighbors}\vec{A}_{ij}\cdot\vec{v}_{ij}.
$$
Restructuring Eq. \ref{eq:doppler_compression_1} results in the Doppler term for $\alpha_i \geq 0$:
\begin{equation}
\label{eq:doppler_term_compression}
    \begin{split}
            &\mathcal{D}_{g,i}[\mathbf{E}_i]=\\
            &\Bigg[\alpha\nu_{g-1/2}\left(\frac{1-R_{2,g-1,i}}{2}\right)\left[-\frac{1}{\Delta_{g-1}}+\frac{\phi\left(r_{g-1,i}\right)}{2\Delta\nu_{g-1}}\right]\Bigg]E_{g-1,i} \\
            &\Bigg[\alpha\nu_{g+1/2}\left(\frac{1-R_{2,g,i}}{2}\right)\left[\frac{1}{\Delta_{g}}-\frac{\phi\left(r_{g,i}\right)}{2\Delta\nu_{g}}\right] \\
            &-\alpha\nu_{g-1/2}\left(\frac{1-R_{2,g-1,i}}{2}\right)\frac{\Delta_{g-1}\phi\left(r_{g-1,i}\right)}{2\Delta\nu_{g-1}\Delta_{g}}\Bigg]E_{g,i}+\\
            &\left[\alpha\nu_{g+1/2}\left(\frac{1-R_{2,g,i}}{2}\right)\frac{\Delta_{g}\phi\left(r_{g,i}\right)}{2\Delta\nu_{g}\Delta_{g+1}}\right]E_{g+1,i},
    \end{split}
\end{equation}
were $\mathbf{E}_i$ is the vector $\mathbf{E}_i\equiv (E_{1,i}\ldots E_{G,i})$, and  $\mathcal{D}_{g,i}[\mathbf{E}_i]$ is the discrete Doppler operator.

Conversely, for $\alpha < 0, \nabla\cdot v > 0$, the energy groups get red-shifted. Thus, the upwind direction is towards the higher energies, and  $E_{\nu_{g\pm 1/2}}$ are reconstructed by interpolation from the right side:   
\begin{equation}
    \begin{split}
        & E_{\nu_g+1/2} = \frac{E_{g+1}}{\Delta_{g+1}}-\frac{\Delta_{g+1}\phi\left(r_{g+1}\right)}{2\Delta\nu_{g+1}}\left(\frac{E_{g+2}}{\Delta_{g+2}}-\frac{E_{g+1}}{\Delta_{g+1}}\right) \\
        & E_{\nu_g-1/2} = \frac{E_{g}}{\Delta_{g}}-\frac{\Delta_{g}\phi\left(r_{g}\right)}{2\Delta\nu_{g}}\left(\frac{E_{g+1}}{\Delta_{g+1}}-\frac{E_{g}}{\Delta_{g}}\right).
    \end{split}
\end{equation}
The same procedure results in the Doppler term for $\alpha_i<0$,
\begin{equation}
 \label{eq:doppler_term_expansion}
    \begin{split}
        &\mathcal{D}_{g,i}[\mathbf{E}_i] = \\
        &\alpha\nu_{g-1/2}\left(\frac{1-R_{2,g-1,i}}{2}\right)\left[-\frac{1}{\Delta_{g}}+\frac{\phi\left(r_{g,i}\right)}{2\Delta\nu_{g}}\right]E_{g,i} \\
        &\Bigg[\alpha\nu_{g+1/2}\left(\frac{1-R_{2,g,i}}{2}\right)\left(\frac{1}{\Delta_{g+1}}+\frac{\phi\left(r_{g+1,i}\right)}{2\Delta\nu_{g+1}}\right)\\
        &-\alpha\nu_{g-1/2}\left(\frac{1-R_{2,g-1,i}}{2}\right)\frac{\Delta_{g}\phi\left(r_{g,i}\right)}{2\Delta\nu_{g}\Delta_{g+1}}\Bigg]E_{g+1,i} \\
        &-\alpha\nu_{g+1/2}\left(\frac{1-R_{2,g,i}}{2}\right)\frac{\Delta_{g+1}\phi\left(r_{g+1,i}\right)}{2\Delta\nu_{g+1}\Delta_{g+2}}E_{g+2,i}.
    \end{split}
\end{equation}
At the frequency boundaries 
$\nu_{1/2}, \nu_{G+ 1/2}$, we impose zero frequency‑space flux, i.e., the Doppler flux across each boundary vanishes.  Inserting all terms into Eq. \ref{eq:multigroup},  and integrating over a single time-step yields
\begin{equation} \label{eq:mg_disc}
        \begin{split}
        &V_{i}E_{g,i}^{n+1}-V_{i}E_{g,i}^{n}=\\
        & \Delta t\Bigl[
        -\left(\frac{1-R_{2,g,i}}{2}\right)E_{g,i}^{n+1}\sum_{j}^\text{neighbors}\vec{A}_{ij}\cdot\vec{v}_{ij} \\
        &+\sum_{j}^\text{neighbors}D_{g,ij}\frac{\vec{A}_{ij}\cdot\left(\overrightarrow {CM}_{j}-\overrightarrow {CM}_{i}\right)}{\left\Vert\overrightarrow {CM}_{j}-\overrightarrow {CM}_{i}\right\Vert^{2}}\left(E_{g,j}^{n+1}-E_{g,i}^{n+1}\right)\\
        & +cV_{i}\kappa_{P, g,i}\left(b_{g,i}U_{m,i}^{n+1}-E_{g,i}^{n+1}\right) - \mathcal{D}_{g,i}[\mathbf{E}_i^{n+1}]\Bigr].
    \end{split}
\end{equation}
Here the superscript $n+1$ denotes values evaluated at the end of the radiation step (in the implicit scheme); quantities without a time index are evaluated at the start of the radiation step. In the Doppler term, all quantities related to the slope limiter are evaluated explicitly.

Equation \ref{eq:mg_disc} is solved together with the material internal energy density equation 
\begin{equation}
    \label{eq:dUmdt}
    \overline{c}_{v}\frac{\partial U_{m}}{\partial t}=c\left(\sum_{g'}\kappa_{P, g'}E_{g'}-\kappa_{P}U_{m}\right),
\end{equation}
where $\overline{c}_{v}\equiv\frac{\partial T}{\partial U_{m}}\frac{\partial u_{m}}{\partial T}$ is the ratio between the heat capacity of the material and the radiation (at the material temperature). Integrating Eq. \ref{eq:dUmdt} over a single time-step results in
\begin{equation}
    \label{eq:dUmdt_implicit}
    \overline{c}_{v}\left(U_{m}^{n+1}-U_{m}^{n}\right)=c\Delta t\left(\sum_{g'}\kappa_{P, g',i}E_{g',i}^{n+1}-\kappa_{P,i}U_{m,i}^{n+1}\right).
\end{equation}
Rearranging Eq. \ref{eq:dUmdt_implicit} and introducing the Fleck factor (\citealt{fleck_implicit_1971})
\begin{equation}
     f_{i}\equiv\frac{1}{1+\frac{c\Delta t}{\overline{c}_{v}}\kappa_{P,i}}
\end{equation}
yields
\begin{equation}
    U_{m}^{n+1}=f_{i}U_{m}^{n}+\frac{1-f_{i}}{\kappa_{P,i}}\sum_{g'}\kappa_{P, g',i}E_{g',i}^{n+1}.
    \label{eq:Um_end}
\end{equation}
Inserting the implicit estimation of $U_m$ (Eq. \ref{eq:Um_end}) into the radiation transport equation (Eq. \ref{eq:mg_disc}) and moving all the implicit terms to the left side of the equality results in the following linear system:
\begin{equation}
    \label{eq:system_mg}
    \begin{split}
        &V_{i}\left(1+c\Delta t\kappa_{P, g,i}\right)E_{g,i}^{n+1}\\
        &+\Delta t\left(\frac{1-R_{2,g,i}}{2}\right)E_{g,i}^{n+1}\sum_{j}\vec{A}_{ij}\cdot\vec{v}_{ij}\\ 
        &-\Delta t\sum_{j}D_{g,ij}\frac{\vec{A}_{ij}\cdot\left(\overrightarrow {CM}_{j}-\overrightarrow {CM}_{i}\right)}{\left\Vert\overrightarrow {CM}_{j}-\overrightarrow {CM}_{i}\right\Vert^{2}}\left(E_{g,j}^{n+1}-E_{g,i}^{n+1}\right) \\
        &-c\Delta tV_{i}\kappa_{P, g,i}b_{g,i}\frac{1-f_{i}}{\kappa_{P,i}}\sum_{g'}\kappa_{P, g',i}E_{g',i}^{n+1}+\Delta t\mathcal{D}_{g,i}[\mathbf{E}_i^{n+1}] \\
        &=V_{i}E_{g,i}^{n}+c\Delta tV_{i}\kappa_{P, g,i}b_{g,i}f_{i}U_{m}^{n}.
    \end{split}
\end{equation}

The final form of the material internal energy equation is given by inserting Eq.~\ref{eq:Um_end} into Eq.~\ref{eq:internal_energy_mg}, resulting in
\begin{equation}
    \label{eq:final_energy}
    \frac{\partial u_{m,i}}{\partial t}=f_ic\left(\sum_{g'}\kappa_{P, g',i}E_{g',i}^{n+1}-\kappa_{P,i}U_{m,i}^n\right).
\end{equation}
Once \eqref{eq:system_mg} is solved and the radiation field at the end of the time step is obtained, we update the internal energy via equation \eqref{eq:final_energy}. The momentum term is updated via
\begin{equation}
        \frac{\M_i^{n+1} - \M^n_{i}}{\Delta t} = -\sum_{g'}\lambda_{g',i}\sum_j^{neighbors} \vec{A}_{ij}\left(\frac{E_{g',i}^{n+1} + E_{g',j}^{n+1}}{2}\right)
\end{equation}
where $\M_i = m_i\vec{v}_i$ is the cell's momentum and $m_i$ is the cell mass. Then the kinetic energy is determined 
\begin{equation}
    V_i K_i^{n+1} = \frac{\left(\M_i^{n+1}\right)^2}{2m_i}
\end{equation}
 And finally, the total cell energy is calculated 
\begin{equation}
    E_{m,i}^{n+1} = u_{m,i}^{n+1} + K_i^{n+1}.
\end{equation}

\subsection{Numerical Solution of the Multigroup Equation}

Equation \ref{eq:system_mg} defines a non-symmetric sparse linear system with $N_{\rm cells}\times N_{\rm groups}$ unknowns. \textsc{rich} solves this linear system using an MPI-parallel, Jacobi-Preconditioned, Bi-Conjugate-Gradient-Stabilized method (BiCGSTAB), an iterative algorithm for solving non-symmetric linear systems (\citealt{BiCGSTAB}).  We solve the sparse linear system \(\mathbf{A}\mathbf{E}=\mathbf{b}\), where \(\mathbf{E}\) stacks the cell–group unknowns. Iterations terminate once all of the following hold:
\begin{enumerate}
\item \textbf{Relative residual:} define \(\mathbf{r}\equiv \mathbf{b}-\mathbf{A}\mathbf{E}\) and require
\[
\|\mathbf{r}\| \le \varepsilon\,\|\mathbf{b}\| .
\]
\item \textbf{Positivity floor:} with \(E_{\max}\equiv \max_j E_j\), enforce for all cells \(i\) and groups \(g\),
\[
E_{g,i} \ge -10^{-10}\,E_{\max}.
\]
\item \textbf{Component-wise residual:} for every unknown \(j\),
\[
|r_j| \le 10^{-6}\,\bigl(|A_{jj}|\,(E_j+10^{-5}\,E_{\max})\bigr).
\]
\end{enumerate}
\subsubsection{Enforcing Energy Conservation}

After the final linear solve, we apply a conservative defect–correction based on the component-wise residual. For an unknown \(j\) representing a cell–group energy density \(E_j\) with cell volume \(V_j\),
the balance defect in energy is \(r_j\).
We then update
\begin{equation}
  V_j E_j^{\mathrm{new}} \leftarrow V_j E_j^{\mathrm{old}} + r_j
  \quad \text{i.e.}\quad
  E_j^{\mathrm{new}} \leftarrow E_j^{\mathrm{old}} + \frac{r_j}{V_j} .
\end{equation}
This post-processing step restores conservation to machine precision without additional iterations. 

\subsubsection{Convergence Acceleration}\label{sec:convergence}
In optically thick cells the radiation–material coupling becomes stiff and the block off-diagonals $c\Delta t\,\kappa_{P, g'}$ increase the condition number, slowing BiCGSTAB. 
In order accelerate the convergence of the BiCGSTAB algorithm, we limit the absorption coefficients to be 

$$\kappa_{P, g}  \leftarrow \min\left(\kappa_{P, g},\,\frac{N_\tau}{c\Delta t}\right)
$$

where $N_\tau$ is the optical depth threshold, which we typically set $N_\tau=100$, which limits the local equilibration time to be $\Delta t/N_\tau$ while leaving the diffusive transport coefficient $\kappa_{R,g}$ unchanged. In our tests, this reduces the iteration count substantially with negligible impact on the physical solution.

\section{Numerical Results} \label{sec:numerical_Results}

\subsection{Marshak Boundary Condition}

The Marshak boundary condition describes the incoming flux into a medium, which is coupled to a heat bath at LTE with temperature $T_{bath}$. LTE means the radiation has a Planck spectrum with the same temperature as the material temperature. The boundary condition is given by  \citep{krief_self-similar_2024,krief2024unified,derei_non-equilibrium_2024}
\begin{equation}
    \label{eq:marshak_boudnary}
    \frac4c F_{inc,g} = E_{g,b} + \frac2c F_{g,b} 
\end{equation}
here $E_{g,b}$ is the group energy at the boundary, $F_{g,b}$ is the total flux on the boundary and 

\begin{equation}
    F_{inc,g} = b_g(T_{bath})\frac{ac}{4}T_{bath}^4 = \frac{c}4U_{bath,g}
\end{equation}
is the incoming group flux from the heat bath and $U_{bath,g}=b_g(T_{bath})aT^4_{bath}$. Assuming cell $i$ is a boundary cell, approximating $E_{g,b}\simeq E_{g,i}$ ,  inserting into Eq. \ref{eq:marshak_boudnary} and rearranging gives

\begin{equation}
    \label{eq:flux_marshak}
    F_{g,b} = \frac c2 \left(E_{g,i} - U_{bath,g}\right)
\end{equation}
In our implementation we use the implicit form of Eq. \ref{eq:flux_marshak}. To the system described in Eq. \ref{eq:system_mg}, for a boundary cell we add $\frac{c\Delta tA_{ij}}{2}E_{g,i}^{n+1}$ , $\frac{c\Delta t}{2}A_{ij}U_{bath,g}$ to the left and right side of the equation, respectively.

\subsection{Grey Diffusion}
\label{sec:grey_validation}
In the optically thick limit (i.e. small mean free path and gradient scale, $R_g\rightarrow 0$ and $\lambda_g\to 1/3$), when taking $\kappa_{P,g}$ and $\kappa_{R,g}$ to be constant, we can recover the grey radiation-diffusion equation by summing the multigroup equations over the energy groups

\begin{equation}    
\frac{\partial E_{r}}{\partial t} +\nabla\cdot\left(\vec{v}E_r\right) =-\frac{E_{r}}{3}\nabla\cdot\vec{v}+\nabla\cdot\left(D\nabla E_{r}\right)+c\kappa_{P}\left(U_{m}-E_{r}\right)
\end{equation}

where $E_r=\sum_g E_g$. We compare the multigroup solution in this degenerate (gray) case to known analytic and semi‑analytic solutions of the grey radiation-diffusion equation. In these cases, the exact group placement is irrelevant, and only adequate coverage of the Planckian between $\nu_{ 1/2},\nu_{G+ 1/2}$ is required. 
\subsubsection{Non-Equilibrium Nonlinear Marshak Waves}

We first test our multigroup diffusion implementation against non-equilibrium, nonlinear supersonic Marshak wave problems (\cite{krief_self-similar_2024,derei_non-equilibrium_2024}). These solutions are self-similar, one-dimensional and radiation-only test problems in the gray diffusion limit. We note that self-similarity methods have recently been used to extend and develop new solutions for radiation transfer (\cite{krief2021analytic,krief2024unified,heizler2024accurate}) and shock hydrodynamics (\cite{giron2021solutions,giron2023solutions,krief2023piston}).

In each non-equilibrium Marshak wave problem, a Marshak boundary condition drives an incoming flux incoming from a heat bath with a prescribed time dependent temperature $T_{bath}\left(t\right)$, into an initially cold material $T_m(x,0)=0$ and a radiation field $E_{r}(x,t)\equiv 0$, with a spatially dependent density profile $\rho(x) = \rho_0 x^{-\omega}$. The radiation temperature is

$$
T_{rad} = \left(\frac{E_{r}}{a}\right)^\frac{1}{4}.
$$

We consider four configurations that differ by opacity scalings and density profiles, and compare them with the analytic solutions at $t=1$ ns. Since these benchmarks are gray solutions, the opacities do not depend on frequency and are only functions of temperature and density. We note that in these benchmarks the Planck (absorption) opacity is much smaller than the Rosseland (total) opacity, so that the medium is optically thick but thermal equilibrium is not reached, while the diffusion limit is reached, so that these solutions are essentially exact transport solutions with high photon scattering and low absorption (see \cite{krief_self-similar_2024,derei_non-equilibrium_2024}). Regarding the numerical setup - all the cases in this section are defined on a uniform grid of 512 cells between $x\in[0,1]$ (with the exception of the last benchmark that uses 512 cells unevenly spaced), using 12 groups spanning $10^{-8}\;\text{keV}$ to  $31\;\text {keV}$ logarithmically. We note that in order to test our multigroup model, we perform multigroup calculations, even though the opacities are frequency independent. The analytic similarity profiles and heat front positions are given as simple closed form analytic formulas in \cite{krief_self-similar_2024} and \cite{derei_non-equilibrium_2024}, respectively.

\subsubsubsection{Problem 1 (Test 2 from \cite{krief_self-similar_2024})}

In this benchmark, the opacities are set to be
\begin{equation}
\begin{split}
&\kappa_{R}\left(T\right)=100\left(\frac{T}{\text{keV}}\right)^{-3}\text{cm}^{-1} \\
&\ \kappa_{P}\left(T\right)=0.001 \kappa_{R}\left(T\right)
\end{split}
\end{equation}

The material energy density and the density profile are uniform and given by:
\begin{equation}
    u\left(T\right)=6.860085\times10^{14}\left(\frac{T}{\text{keV}}\right)^{4}\ \text{erg}\ \text{cm}^{-3}
\end{equation}

$$
\rho(x)=1\ \text{g}\ \text{cm}^{-3}
$$

and the driving bath temperature is

\begin{equation}
    T_{bath}\left(t\right)=1.008038\left(\frac{t}{\text{ns}}\right)^{\frac{1}{3}}\text{keV}.
\end{equation}

\begin{figure*}
    \centering
    \begin{minipage}{0.49\textwidth}
    \includegraphics[width=1.0\linewidth]{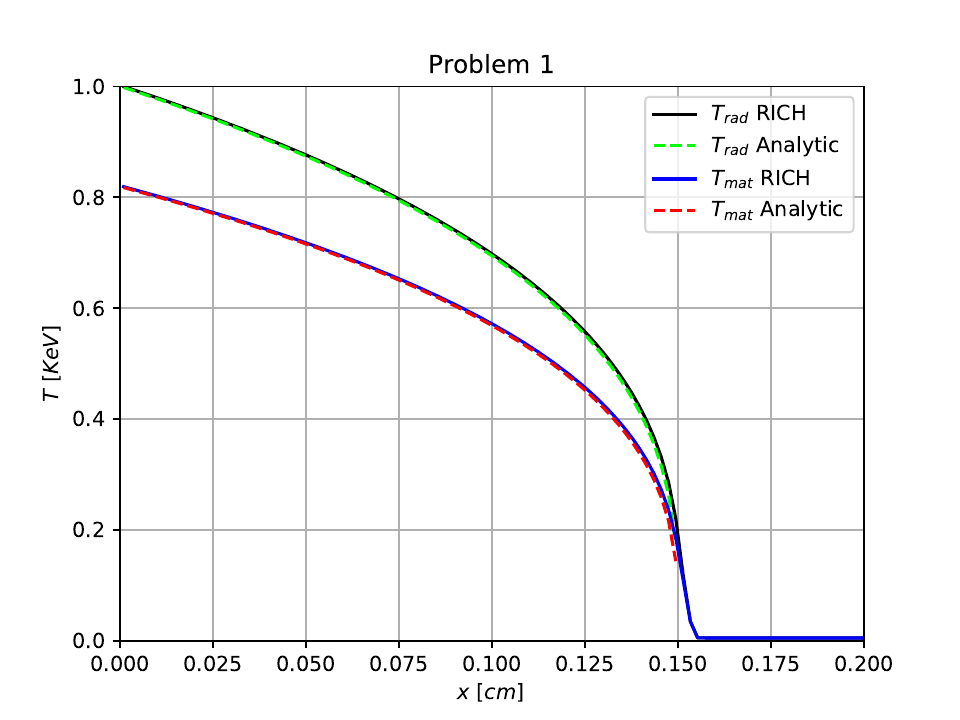}
    \end{minipage}
    \begin{minipage}{0.49\textwidth}
    \includegraphics[width=1.0\linewidth]{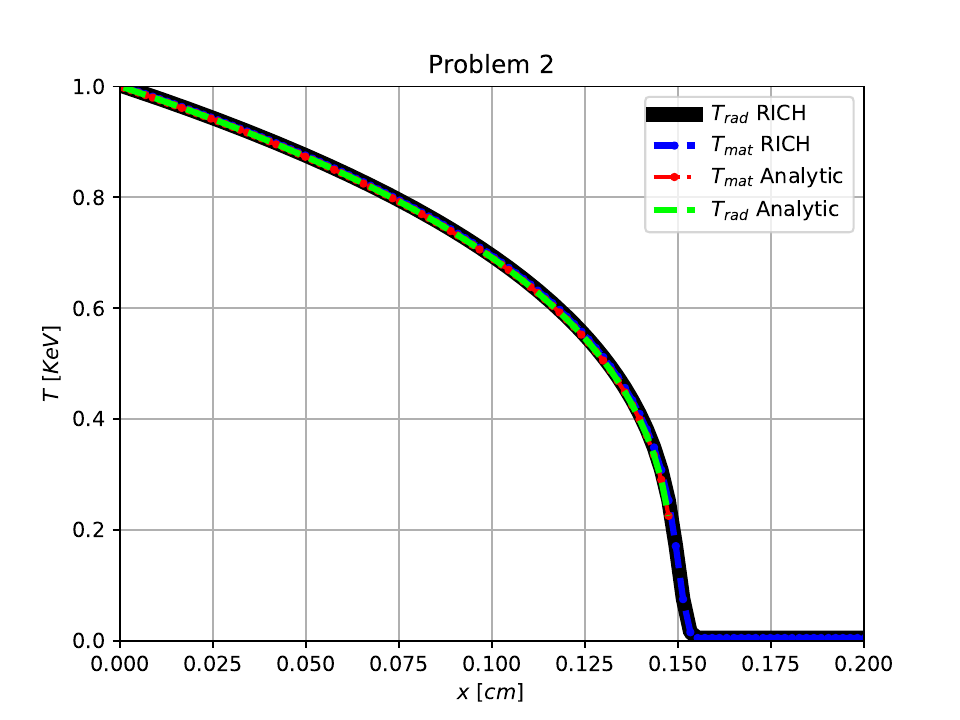}
    \end{minipage}
    \caption{Comparison of the radiation temperature $T_{\rm rad}$ and material temperature $T_{\rm mat}$ at $t=1\ \text{ns}$ between the \textsc{rich} simulation and the analytic solution for Marshak problems 1 (left) and 2 (right). In problem 2 the absorption coefficient is large enough to enforce LTE, so $T_{\rm rad}\simeq T_{\rm mat}$.}
    \label{fig:Marshak_problem12}
\end{figure*}

\subsubsubsection{Problem 2 (Test 3 from \cite{krief_self-similar_2024})}

This problem is the equilbrium limit of the previous benchmark. The opacities are now set to be
\begin{equation}
    \kappa_{R}\left(T\right)=\kappa_{P}\left(T\right)=100\left(\frac{T}{\text{keV}}\right)^{-3}\text{cm}^{-1}.
\end{equation}

The material energy density and the density profile are
\begin{equation}
    u\left(T\right)=6.860085\times10^{14}\left(\frac{T}{\text{keV}}\right)^{4}\ \text{erg}\ \text{cm}^{-3},\quad 
\end{equation}

$$
\rho(x)=1\ \text{g}\ \text{cm}^{-3}
$$

and the bath temperature is

\begin{equation}
    T_{bath}\left(t\right)=1.014565 \left(\frac{t}{\text{ns}}\right)^{\frac{1}{3}}\text{keV}.
\end{equation}
Figure \ref{fig:Marshak_problem12} shows the material and radiation temperature as function of distance inside the slab for both benchmarks (1 \& 2), and they agree nicely with the analytic solution.

\subsubsubsection{Problem 3 (Test 1 from \cite{derei_non-equilibrium_2024})}

In this benchmark the opacities are
\begin{equation}
\begin{split}
&\kappa_{R}\left(T,\rho\right)=40\left(\frac{T}{\text{keV}}\right)^{-1.5}\left(\frac{\rho}{\text{g}\ \text{cm}^{-3}}\right)^{1.2}\text{cm}^{-1} \\
&\ \kappa_{P}\left(T,\rho\right)=0.0025\kappa_{R}\left(T,\rho\right)
\end{split}
\end{equation}

The material energy density and the density profile are spatially non-uniform and given by
\begin{equation}
    u\left(T,\rho\right)=10^{14}\left(\frac{T}{\text{keV}}\right)^{3.4}\left(\frac{\rho}{\text{g}\ \text{cm}^{-3}}\right)^{0.86}\text{erg}\ \text{cm}^{-3} \\
\end{equation}

$$
    \rho(x)=\left(\frac{x}{\text{cm}}\right)^{\frac{20}{19}}\ \text{g}\ \text{cm}^{-3}
$$
and the bath temperature is

\begin{equation}
    T_{bath}\left(t\right)=1.0470478 \left(\frac{t}{\text{ns}}\right)^{\frac{86}{57}}\text{keV}
\end{equation}

\subsubsubsection{Problem 4 (Test 3 from \cite{derei_non-equilibrium_2024})}

The opacities are now set to be
\begin{equation}
\begin{split} 
&\kappa_{R}\left(T,\rho\right)=2\left(\frac{T}{\text{keV}}\right)^{-4.5}\left(\frac{\rho}{\text{g}\ \text{cm}^{-3}}\right)^{1.9}\text{cm}^{-1} \\
&\kappa_{P}\left(T,\rho\right)=0.0005\kappa_{R}\left(T,\rho\right)
\end{split}
\end{equation}

The material energy density and the density profile are (notice that the density profile diverges at $x=0$)
\begin{equation}
    u\left(T,\rho\right)=10^{14}\left(\frac{T}{\text{keV}}\right)^{6}\left(\frac{\rho}{\text{g}\ \text{cm}^{-3}}\right)^{0.7}\ \text{erg}\ \text{cm}^{-3}
\end{equation}
$$
    \rho(x)=\left(\frac{x}{\text{cm}}\right)^{-\frac{40}{139}}\ \text{g}\ \text{cm}^{-3}
$$
and the bath temperature is

\begin{equation}
    T_{bath}\left(t\right)=1.01008116  \left(\frac{t}{\text{ns}}\right)^{\frac{14}{139}}\text{keV}.
\end{equation}

\begin{figure*}
    \centering
    \begin{minipage}{0.49\textwidth}
    \includegraphics[width=1.0\linewidth]{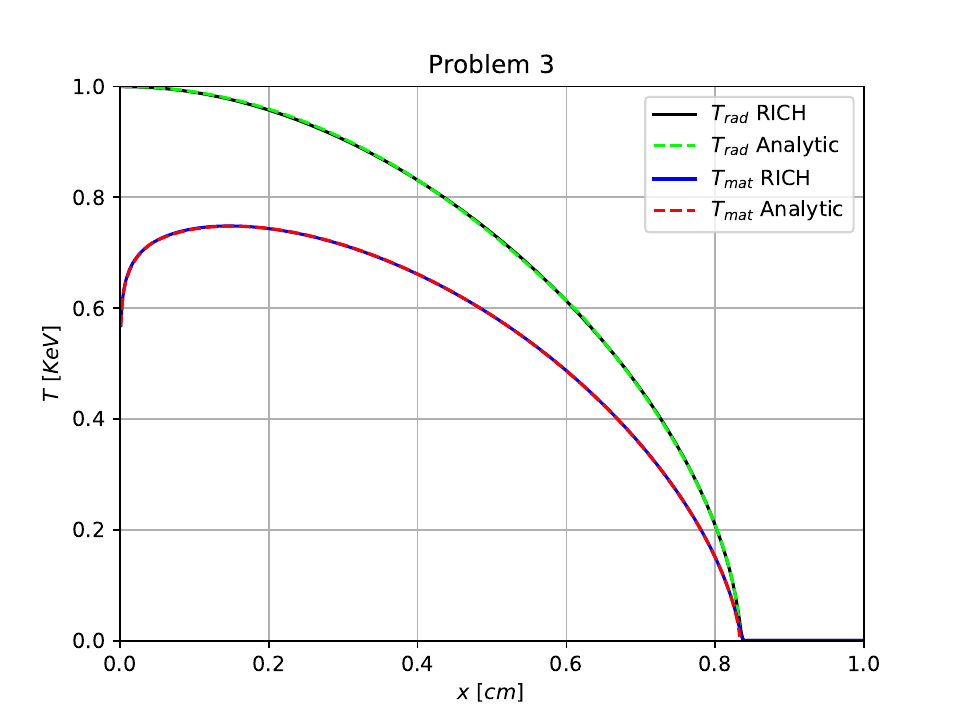}
    \end{minipage}
    \begin{minipage}{0.49\textwidth}
    \includegraphics[width=1.0\linewidth]{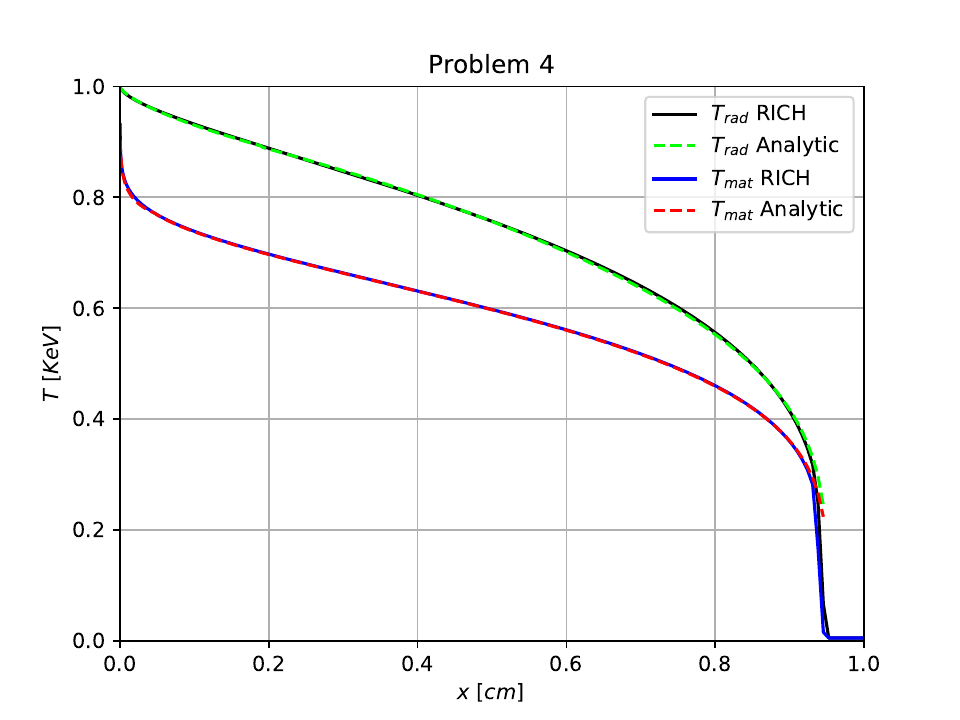}
    \end{minipage}
    \caption{Comparison of the radiation temperature and material temperature at $t=1\ \text{ns}$ between the \textsc{rich} simulation and the analytic solution for Marshak problems 3 (left) and 4 (right).}
    \label{fig:Marshak_problem34}
\end{figure*}

Unlike the previous benchmarks, the divergence of the density at the origin required us to use a refined grid in which the Voronoi mesh generating points are located at 
\begin{equation}
    x_i = 10^{-5} + 2.24\times10^{-2}\left(1.0075^i - 1\right)\ \text{cm}.
\end{equation}
Figure \ref{fig:Marshak_problem34} shows the material and radiation temperature as function of distance inside the slab for both benchmarks (3 \& 4), and they agree nicely with the analytic solution. 

\subsubsection{Mach 2 Radiative Shock}

In order to test our coupling of the radiation transport to the hydrodynamics, we run the Mach 2 test problem \citep{ferguson_nonrelativistic_2017, Lowrie2008}. This is a one-dimensional, non-equilibrium, radiation-hydrodynamics test problem in the grey diffusion approximation. 

The problem consists of two semi-infinite regions of a perfect gas with the same adiabatic index $\gamma$. The problem is defined using the values of the gas at $-\infty$  for the left region $\rho_\ell, T_\ell, v_\ell$ and at $+\infty$ for the right region $\rho_{r},T_{r},v_{r}$. 

A shock front develops between the two regions, and in the shock frame  the steady state solution has a semi-analytic solution.  

 The initial conditions for the left region $x < 0$ are:

\begin{equation}
    \rho_\ell = 5.45887\times 10^{-13}\ \text{g}\ \text{cm}^{-3}
\end{equation}
$$
    T_{\ell} = 100\ \text{K},\ v_{\ell} = 2.3547\times 10^5\ \text{cm}\ \text{s}^{-1}.
$$
And for the right region $x > 0$
\begin{equation}
    \rho_r = 1.2479\times 10^{-12} \text{g}\ \text{cm}^{-3}
\end{equation}
$$ 
    T_{r} = 207.757\ \text{K},\ v_{r} = 1.03\times 10^5 \ \text{cm}\ \text{s}^{-1}.
$$

The properties of the gas are given by

\begin{equation}
    \begin{split}
    &\gamma = \frac{5}{3},\ c_{v} = 1.2472\times 10^8\ \text{erg}\ \text{K}^{-1}\ \text{g}^{-1}\\
    &\kappa_R=0.848902\ \text{cm}^{-1},\ \kappa_P=3.93\times10^{-5} \ \text{cm}^{-1}
    \end{split}
\end{equation}

\begin{figure*}
    \centering
    \begin{minipage}{0.49\textwidth}
    \includegraphics[width=1.0\linewidth]{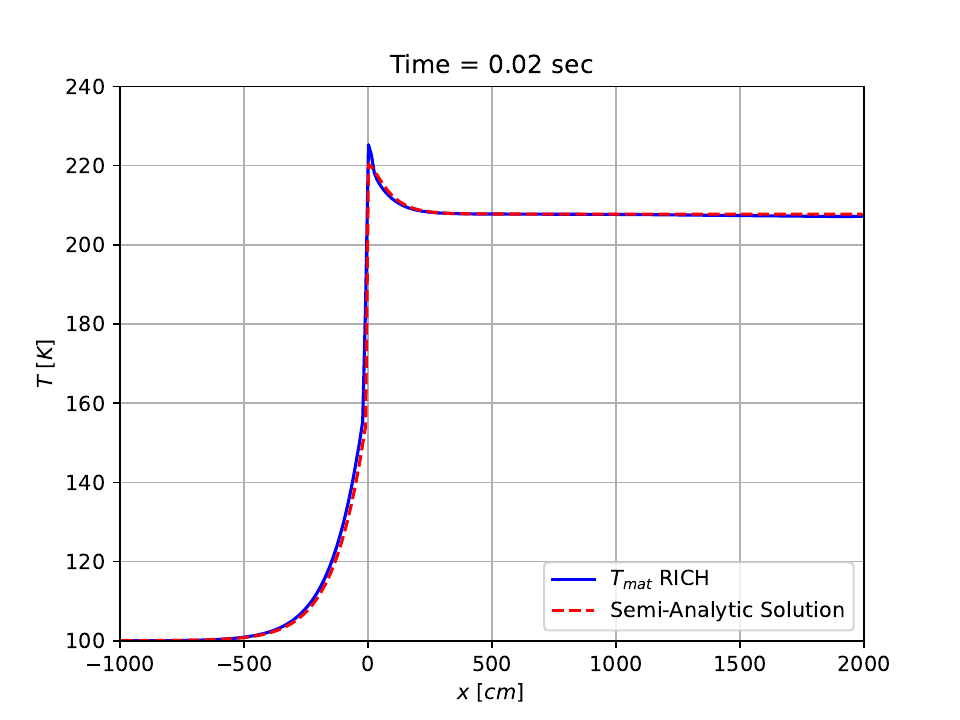} 
    \end{minipage}
    \begin{minipage}{0.49\textwidth}
    \includegraphics[width=1.0\linewidth]{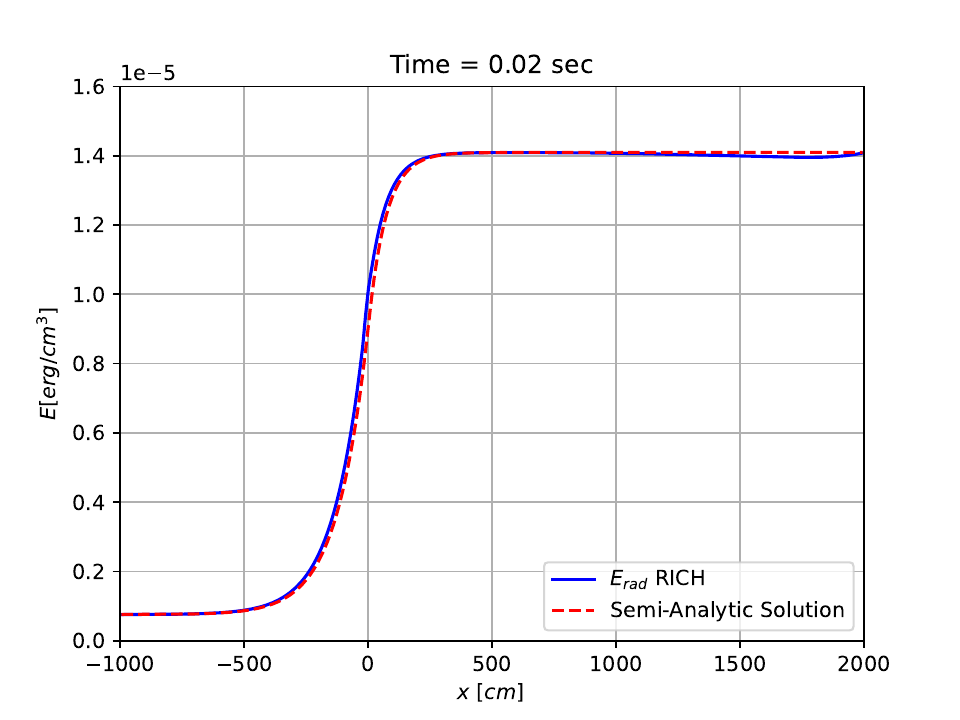}
    \end{minipage}
    \caption{The gas temperature (left) and radiation energy density (right) for the Mach 2 benchmark along with the semi-analytic solution.}
    \label{fig:mach2}
\end{figure*}

Where $c_v = \frac{\partial u_m}{\partial T}$ is the specific heat capacity. The simulation is performed using a uniform grid of 256 cells between $x=-1000$ cm and $x=2000$ cm, with $5$ energy groups between $10^{-3}\text{eV}$ and $1\text{eV}$ spaced logarithmically. We evolve the system until $t=0.02$ seconds, by then the system has evolved to the steady state solution. Figure~\ref{fig:mach2} shows the gas temperature and radiation energy density profiles along with the semi‑analytic solution \footnote{For the Mach 2 radiative problem, we used the analytical solver by Ulrich Noebauer, available in the \texttt{public-astro-tools} repository at \url{https://github.com/unoebauer/public-astro-tools/tree/master}.}
.

\subsection{Multigroup benchmarks}
\label{sec:multigroup_validation}
\subsubsection{Densmore Multigroup tests}

We validate our multigroup diffusion module by comparing it with the benchmarks presented in \citet{DENSMORE20126924}. The benchmarks consists of one-dimensional, radiation-only, non-linear multigroup problems. The reference data from \citet{DENSMORE20126924} (see also \citet{steinberg2022multi,steinberg2023frequency}) were digitized from their figures, which were obtained using a Monte Carlo method that remains reliable in low-opacity regimes.

All of the setups consist of a Marshak boundary condition at the left boundary located at $x=0$, with a bath temperature of $1\;\text{keV}$, an initially cold ideal gas in the domain $x\in[0,L]$, and an open boundary at $x=L$. The parameters of the equation of state are: 

\begin{equation}    
    \gamma = 1.4,\quad c_v = \frac{10^{15}}{\text{keV}}\ \text{erg}\ \text{cm}^{-3}.
\end{equation}
The opacity has the following spatial, temperature and frequency dependence: 
\begin{equation}
    \kappa_{R,g}\left(x, T\right) = \kappa_{P, g}(x,T) = \frac{\kappa_{0}\left(x\right)}{\nu_{g}^{3}\sqrt{k_{B}T}} \text{cm}^{-1}
\end{equation}
and the different benchmarks differ in their corresponding $\kappa_0(x)$. All simulations are run with 32 logarithmically spaced groups between $10^{-4}\;\text{keV}$ and $100\;\text{keV}$ with a uniform grid of 128 cells between  $[0,L]$. 

In the first three runs, $\kappa_0(x)$ was constant and equal to  
\begin{equation}
    \kappa_0 = 10\;\text{keV}^{3.5}, 1000\;\text{keV}^{3.5}, 100\;\text{keV}^{3.5}
\end{equation}
The system length $L$ differs between the runs and is respectively
\begin{equation}
    L = 5,\; 3, \;1\ \text{cm}
\end{equation}

\begin{figure*}
    \centering
    \begin{minipage}{0.49\textwidth}
    \includegraphics[width=1.0\linewidth]{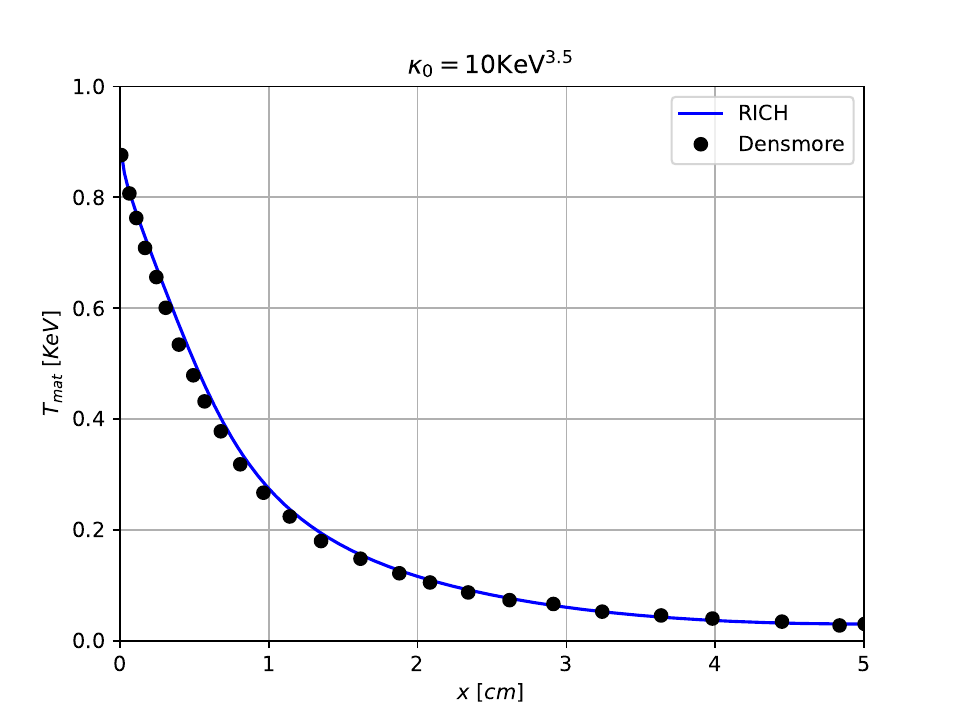}
    \end{minipage}
    \begin{minipage}{0.49\textwidth}
    \includegraphics[width=1.0\linewidth]{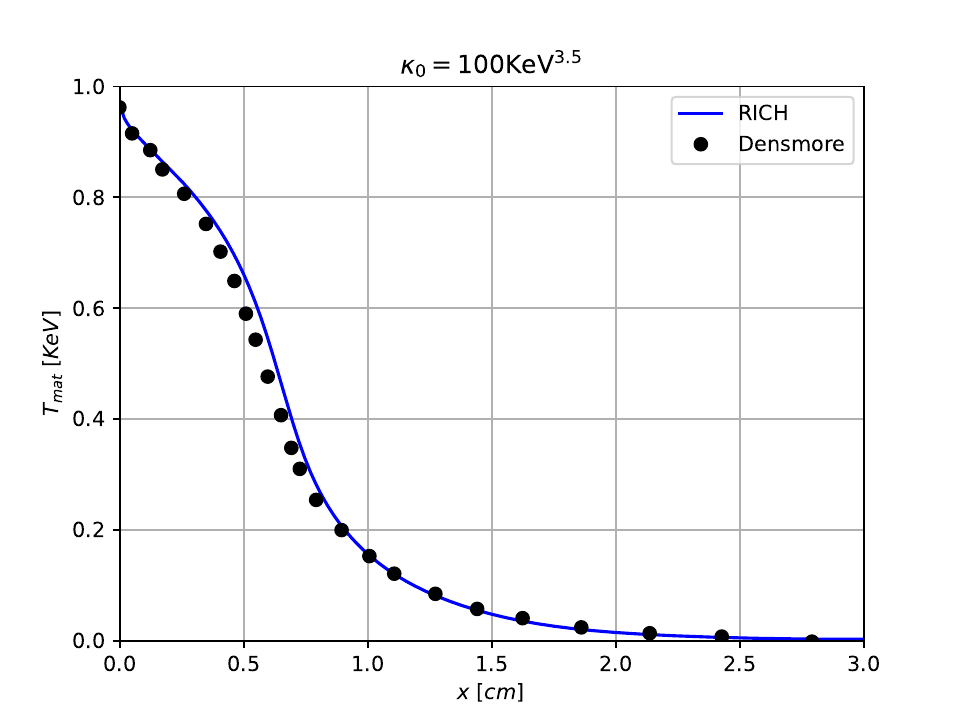}
    \end{minipage}
    \caption{Comparison of the gas temperature at $t=1$ ns between \textsc{rich} and the Monte-Carlo simulation results taken from \citet{DENSMORE20126924} for the cases $\kappa_0 = 10\;\text{keV}^{3.5}$ (left) and $\kappa_0 = 100\;\text{keV}^{3.5}$ (right)}.
    \label{fig:densmore_10kev_100kev}
\end{figure*}

For the fourth run $\kappa_0(x)$ was a step function, jumping from an optically thin region to an optically thick region

\begin{equation}
    \label{eq:densmore_step_function_opacity}
    \kappa_0(x) = 
    \begin{cases}
        10\;\text{keV}^{3.5} & , x < 2 \\
        1000\;\text{keV}^{3.5} & , x \geq 2
    \end{cases}
\end{equation}

with a system of $L=3$ cm. 

\begin{figure*}
    \centering
    \begin{minipage}{0.49\textwidth}
    \includegraphics[width=1.0\linewidth]{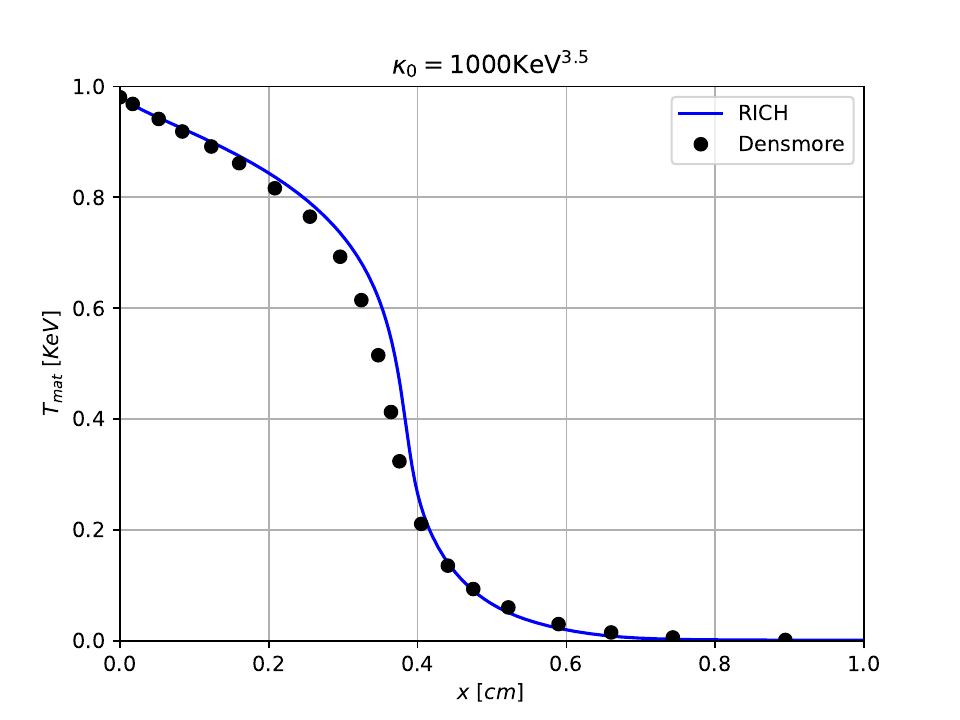}
    \end{minipage}
    \begin{minipage}{0.49\textwidth}
    \includegraphics[width=1.0\linewidth]{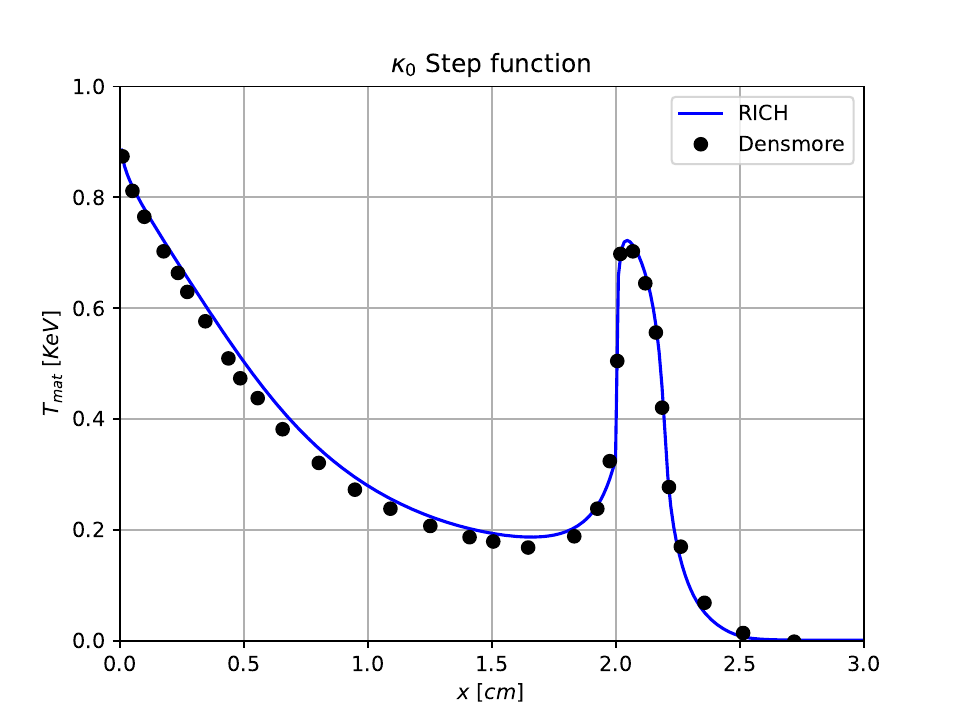}
    \end{minipage}
    \caption{Comparison of the gas temperature at $t=1$ ns between \textsc{rich} and the Monte-Carlo simulation results taken from \citet{DENSMORE20126924} for the medium opacity case $\kappa_0 = 1000\;\text{keV}^{3.5}$ (left) and the step function opacity as presented in Eq.~\ref{eq:densmore_step_function_opacity} (right).}
    \label{fig:densmore_1000kev_step}
\end{figure*}

Figures~\ref{fig:densmore_10kev_100kev} and \ref{fig:densmore_1000kev_step} show the gas temperature at the time $t=1\; \text{ns}$ for the different benchmarks. Overall, even though the benchmarks are compared to a Monte-Carlo method, the agreement is good for all of the tests.

\subsubsection{Doppler Term}\label{sec:doppler_test}

To validate the Doppler term, we construct a zero-dimensional test problem. Starting from Eq.~\ref{eq:multifrequency}, we consider the optically thick limit and omit advection, diffusion, and emission–absorption terms, isolating the Doppler contribution:

\begin{equation}
    \label{eq:Doppler_test_initial}
    \frac{\partial E_{\nu}}{\partial t}=-\left(\frac{1}{3}E_{\nu}-\frac{1}{3}\frac{\partial\left(\nu E_{\nu}\right)}{\partial \nu}\right)\nabla\cdot \vec{v}
\end{equation}
This configuration constitutes an artificial variant of the radiation–diffusion equation. We include the hydrodynamic dependence through $\nabla\cdot\vec{v}$, but intentionally omit it from the advection term in order to isolate the effect of the Doppler contribution.

In order to simplify the notation, we denote $K \equiv -\frac{\nabla\cdot \vec{v}}{3}$ and rewrite Eq.~\eqref{eq:Doppler_test_initial} as:

\begin{equation}
    \frac{1}{K}\frac{\partial E_{\nu}}{\partial t}+\frac{\partial\left(\nu E_{\nu}\right)}{\partial\nu}=E_{\nu}.
\end{equation}

This can be simplified to be:

\begin{equation}
    \frac{1}{K}\frac{\partial E_{\nu}}{\partial t}+\nu\frac{\partial E_{\nu}}{\partial\nu}=0.
    \label{eq:doppler_mid}
\end{equation}
Assuming $K$ is constant, equation~\eqref{eq:doppler_mid} is known to have a solution in the form \citep{polyanin_handbook_2001}

\begin{equation}
    E\left(\nu,t\right)=\Phi\left(\nu e^{-Kt}\right),
\end{equation}
where $\Phi$ is a function that is given by the initial conditions. By equating the two sides at time $t=0$

\begin{equation}
    E(\nu, 0) = \Phi(\nu)    
\end{equation}
we can recover the general solution as a function of frequency and time
\begin{equation}
    E\left(\nu,t\right)=E\left(\nu e^{-Kt},0\right).
\end{equation}

For this benchmark, we set $D_g= 0$, $\kappa_{P,g}= 0$ and $\lambda=\frac{1}{3}$ in \textsc{rich}, and run only the radiation transport (no hydrodynamic evolution). 

The test consists of two identical box-shaped cells. The left cell is stationary with $\vec{v}_{\ell} = \vec{0}$, while the right cell is assigned a velocity of $\vec{v}_{r} = (10^{9}, 0, 0)\;\text{cm}\cdot\text{s}^{-1}$. This configuration generates a non-zero $\nabla\cdot\vec{v}$, with opposite signs in the two cells, thereby creating a controlled expansion in the left cell and compression in the right cell, ideal for testing the Doppler term.

\begin{figure*}[t]
    \centering
    \begin{minipage}{0.49\textwidth}
    \includegraphics[width=1.0\linewidth]{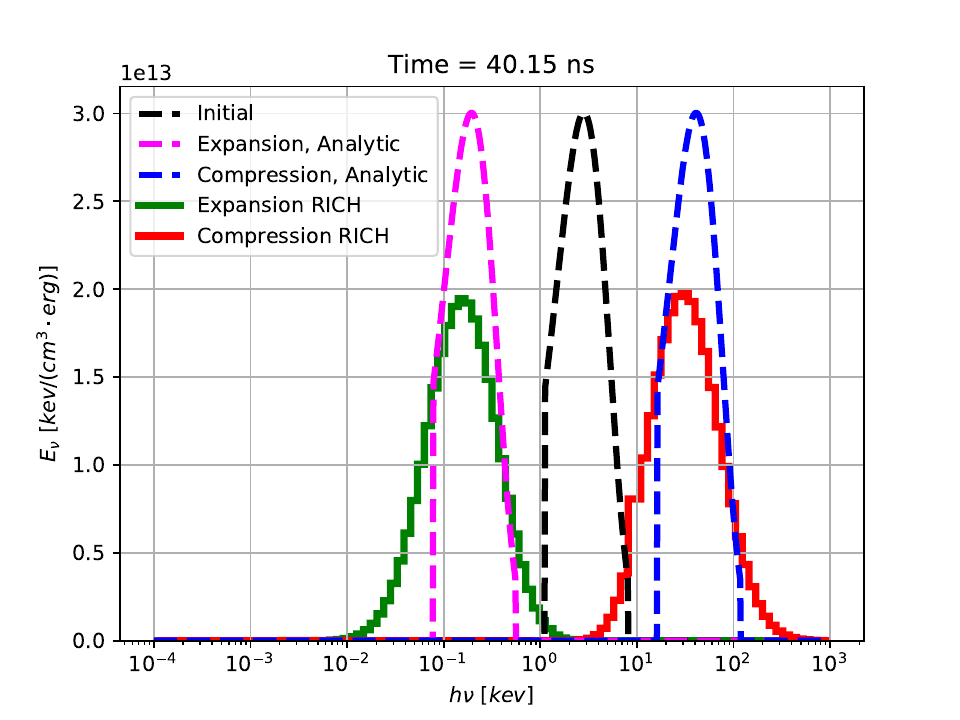} 
    \end{minipage}
    \begin{minipage}{0.49\textwidth}
    \includegraphics[width=1.0\linewidth]{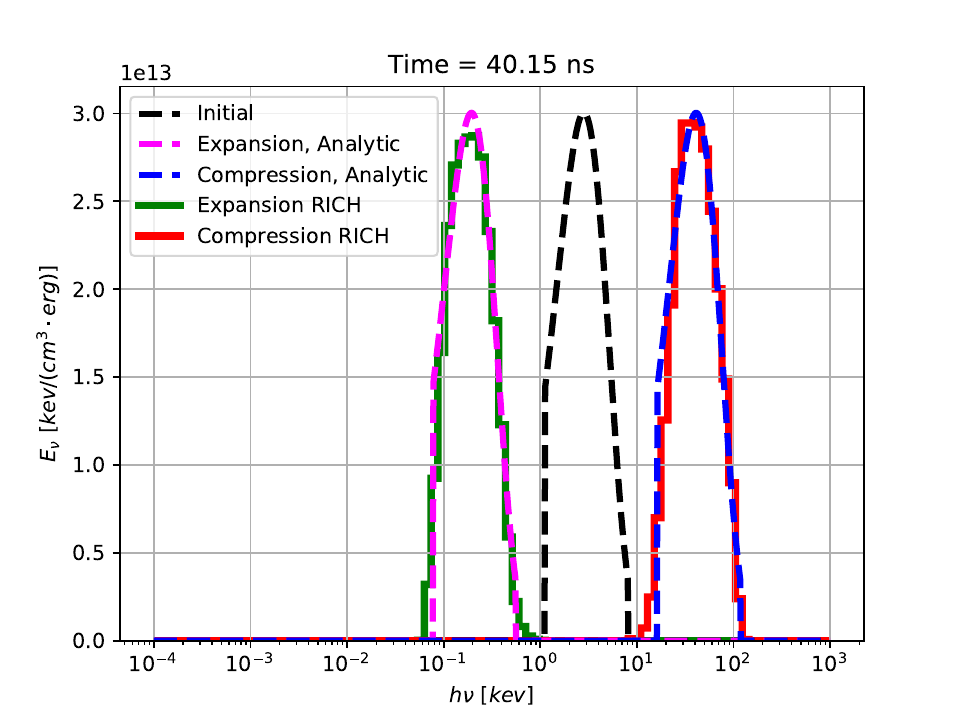}
    \end{minipage}
    \caption{Comparison between the \textsc{rich} and the analytic solution for the case of the upwind method (left) and the Superbee slope-limiter (right) at a time $t=40$ ns. Using the Superbee limiter drastically reduces the numerical diffusion compared to the simple upwind method in both the compression and expansion cases.}
    \label{fig:doppler}
\end{figure*}

The initial radiation profile (which is arbitrary), is given by a truncated Planck function with a temperature of 1 $\text{keV}$
\begin{equation}
    E\left(\nu,0\right)=\begin{cases}
B\left(\nu,T= 1\text{keV}\right) & 1.12\;\text{keV}<\nu<8.12\;\text{keV}\\
0 & \text{else.}
\end{cases}
\end{equation}
To assess the impact of slope limiting (see Eqs.~\ref{eq:doppler_term_compression} and \ref{eq:doppler_term_expansion}), we compare a first-order upwind scheme with $\phi = 0$ to the Superbee limiter (Eq.~\ref{eq:superbee_slope_limiter}). As illustrated in Fig.~\ref{fig:doppler}, the Superbee limiter yields markedly lower numerical diffusion than the upwind approach in both compressive and expansive regimes, resulting in a more accurate solution of the Doppler term.

\subsubsection{Convergence Acceleration\label{sec:convergence_test}}
\begin{figure*}[t]
    \centering
    \begin{minipage}{0.49\textwidth}
    \includegraphics[width=1.0\linewidth]{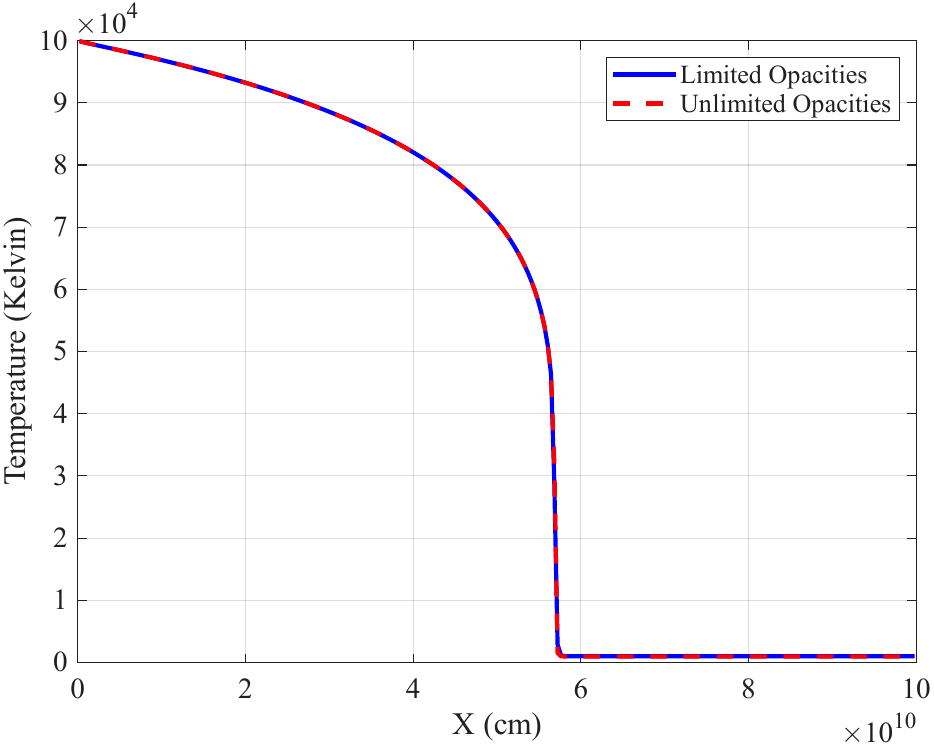} 
    \end{minipage}
    \begin{minipage}{0.49\textwidth}
    \includegraphics[width=1.0\linewidth]{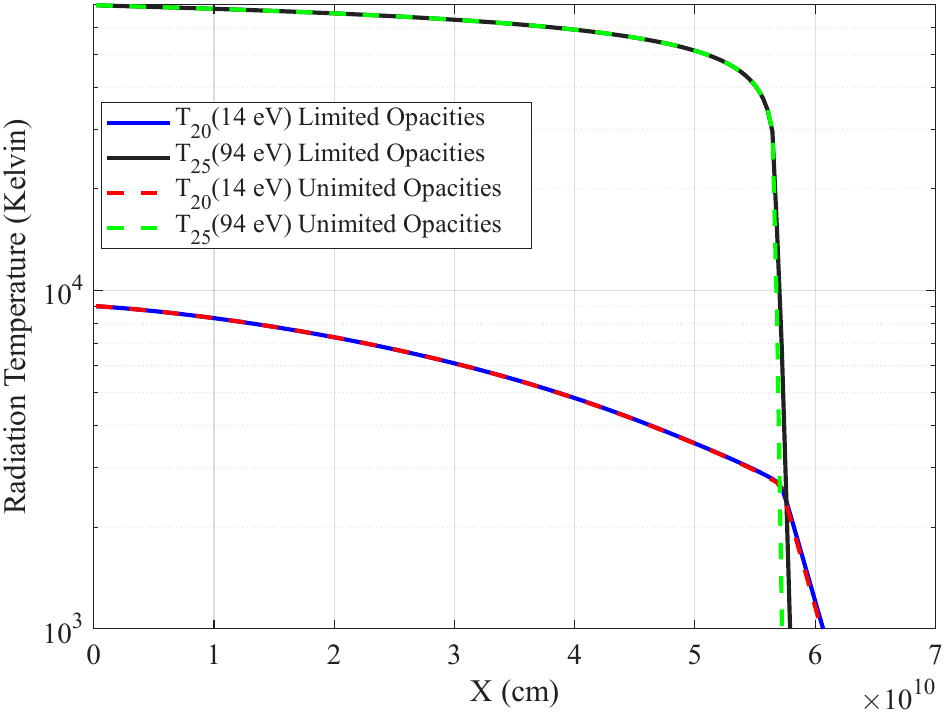}
    \end{minipage}
    \caption{Temperature profiles of the gas (left) and of the radiation (right) in two energy groups (14 eV and 94 eV) for simulations conducted with and without the limiting of the absorption coefficients. The overlapping curves confirm that the method maintains solution accuracy. The accelerated run reduces the computational cost by more than an order of magnitude, completing in 82 s compared to 923 s.}
    \label{fig:convergence}
\end{figure*}

The improved convergence method described in section~\ref{sec:convergence} is evaluated using an astrophysically motivated benchmark. Specifically, we model a Marshak wave under physical conditions characteristic of Tidal Disruption Event simulations \citep{steinberg_streamdisk_2024}. The setup consists of a 1D domain with a Marshak boundary condition imposed at one end, where a bath temperature of $T_{bath}=10^5\;\text{K}$ drives radiation into cold gas of density $\rho=10^{-7}\;\text{g/cc}$. The gas follows an ideal-gas equation of state with adiabatic index $\gamma=5/3$ and a heat capacity of $C_v=10^8\rho\;\text{erg}\;\text{cm}^{-3}\;\text{K}^{-1}$. For the opacity we assume a free-free opacity described by $\kappa_{P,\nu} = \kappa_{R,\nu}=3.7\cdot 10^8n^2\nu^{-3}T^{-1/2}\cdot(1-\exp{(-h\nu/k_BT)})\;\text{cm}^{-1}$, assuming the gas is completely ionized at all times and all of the quantities are in CGS units. We evolve the system until a time of $t=10^3\;\text{s}$ on a uniform grid $x\in(0,\;10^{11})\;\text{cm}$ divided uniformly into 256 cells once with our convergence acceleration using $N_\tau=100$ and once without it. Figure \ref{fig:convergence} presents the gas temperature and the radiation temperatures of two energy groups, $14\ \rm eV$ and $94\ \rm eV$ as functions of position for both simulations. The profiles are nearly identical, confirming that the convergence-acceleration method does not alter the physical solution. However, the computational cost differs substantially: the run without acceleration required 923 s, whereas the accelerated run completed in only 82 s.
\subsection{Tidal Disruption Event}

To demonstrate the astrophysical applicability of the new multigroup radiation–diffusion module, we employ it in a full, three dimensional, self-consistent tidal disruption event (TDE) simulation. We simulate the parabolic approach of a star with mass $M_\star = 0.5 M_\odot$ and radius $R_\star = 0.47R_\odot$ towards a $M_\bullet = 10^4M_\odot$ black hole.  The star is quickly torn apart by the black hole's tidal field, and its deconfined debris flies out to radii $\sim R_\star (M_\bullet / M_\star)^{2/3}$ before the most bound material turns around at its apocenter.  As stellar debris returns to pericenter, it begins dissipating orbital energy into heat through shocks.  Eventually, a subset of this dissipated energy is radiated, producing the luminous nuclear flares seen today by time-domain observers \citep{vanVelzen+20}.  The details of shock dissipation, debris circularization, and radiative emission all remain hotly debated at present \citep{Hayasaki+13, Piran+15, Shiokawa+15, BonnerotStone21, Bonnerot+21, Andalman+22, steinberg_streamdisk_2024, Huang+24, Andalman+25}, and this proof-of-principle simulation does not aim at resolving major open questions in TDE physics.  Because the simulation has the exact astrophysical parameters as the grey-\textsc{rich} simulations of \citep{Martire+25}, we generally expect a similar result: much weaker dissipation and slower circularization than what is seen in supermassive black hole TDEs; emission powered by dissipation at the nozzle shock that is reprocessed through a radiation-driven wind; near-Eddington peak luminosities.  The main advantage of this work relative to past end-to-end TDE simulations is the use of multigroup (rather than grey) radiation transfer, which allows us to self-consistently predict light curve evolution in different frequency bands.

For this simulation, the maximum resolution reached was $18\times10^6$ cells, divided across 256 cores. The gas opacities are interpolated from tabular data calculated using the STAR atomic code, which employs the super-transition-array method \citep{krief2018new,krief2015effect,krief2015variance,krief2018effect}. For the radiation, we used 10 energy groups, presented in Table \ref{tab:TDE_energy_groups}.  Together these groups span the full range of energies in which observed TDEs have been seen to emit quasi-thermally\footnote{I.e. we ignore the radio and gamma-ray photons that are produced via non-thermal processes related to outflows in a subset of TDEs.}.    
\begin{table}
    \centering
    \begin{tabular}{|c|c|c|}\hline
         Name&  $\nu_{low}\ [eV]$& $\nu_{high}\ [eV]$\\\hline
         i&  0.01& 1.6767\\\hline
         r&  1.6767& 2.4796\\\hline
         b&  2.4796& 3.3508
\\\hline
         U&  3.3508
& 4.1327\\\hline
         UVW2&  4.1327& 7.7488\\\hline
         H&  7.7488& 15\\\hline
         He&  15& 100\\\hline
         Super Soft (SS)&  100& 300\\\hline
         Soft X-ray&  300& 1000\\ \hline
 Hard X-ray& 1000&20000\\\hline
    \end{tabular}
    \caption{Photon energy groups used in the multigroup TDE simulation.  Most thermal emission from TDEs has been observed in our r-, b-, U-, UVW2, SS- and Soft X-ray bands.  Theoretical and observational arguments suggest that  the majority of TDE emission may lie in unobservable EUV wavelengths, parametrized here as ``H-'' and ``He-'' bands. }
    \label{tab:TDE_energy_groups}
\end{table}
A more detailed description of the setup, initial conditions, and adaptive mesh refinement criteria is given in \citet{Martire+25}. The simulation is run until it reaches a time of 2.1 fallback times ($t_{\rm fb} = 2.57\ \text{days}$), at which point it has achieved its peak bolometric luminosity.

\begin{figure}
    \centering
    \includegraphics[width=1.0\linewidth]{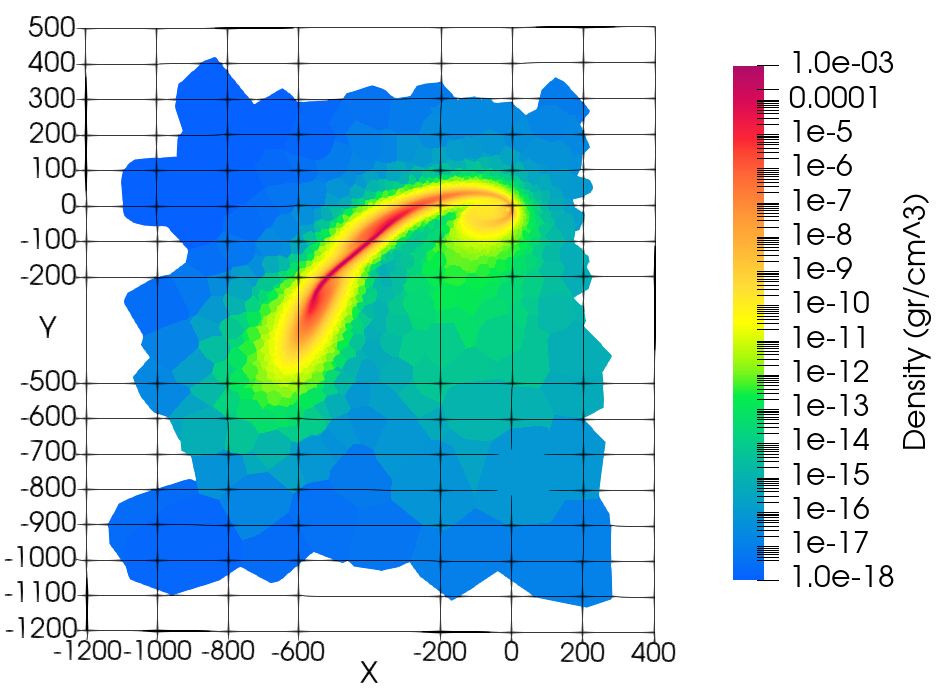}
    \caption{Mass density slice of the TDE simulation in the orbital plane at $t=0.42\ t_{\rm fb}\approx 1\ \text{day}$ after the star was disrupted. The stellar debris has begun falling back, forming an accretion flow around the black hole.  Gas densities are labeled according to the color bar in the figure, and are highest in the debris stream. The black hole is placed at the origin, $X,Y$ are in $R_\odot$.}
    \label{fig:TDE_density}
\end{figure}

Figure \ref{fig:TDE_density} shows a density slice through the orbital plane of the simulation at $t=0.42\ t_{\rm fb}$.  At this point, the most tightly bound debris has just begun to return to pericenter, and dissipation in the nozzle shock is beginning to produce the rising phase of the light curve.

We extract the luminosity in each energy group along a given viewing angle $\vec{\Omega}$ (the angular grid is generated using HEALPix with $N_{\rm side} = 8$, \citealt{2005ApJ...622..759G}) 
\begin{equation}
    L_{g}(r_{\vec{\Omega}}, \vec{\Omega}, t) = 4\pi r_{\vec{\Omega}}^2cE_g\left(r_{\vec{\Omega}}, \vec{\Omega}, t\right)
\end{equation}
by identifying the distance $r_{\vec{\Omega}}$ along a ray originating at the black hole, where the outward group optical depth reaches
\begin{equation}
\tau_{\rm g} = \intop_{r_\Omega}^\infty \kappa_{R,g} dr = \frac{2}{3}.
\end{equation}

\begin{figure*}
    \begin{minipage}{0.49\textwidth}
    \includegraphics[width=1.0\linewidth]{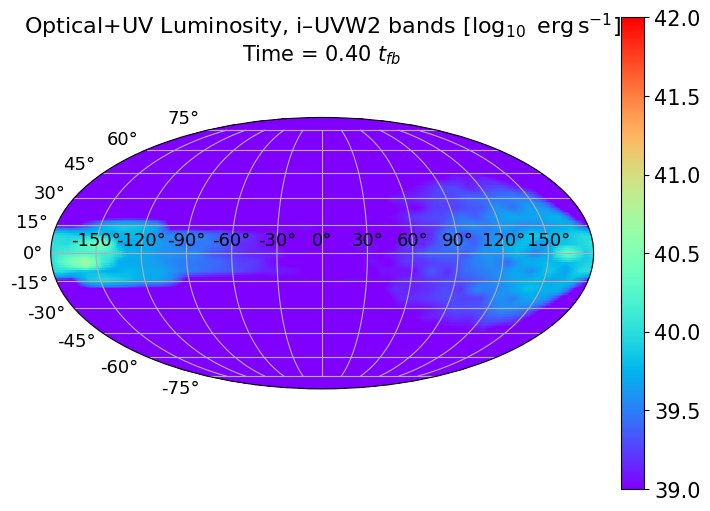} 
    \end{minipage}
    \begin{minipage}{0.49\textwidth}
    \includegraphics[width=1.0\linewidth]{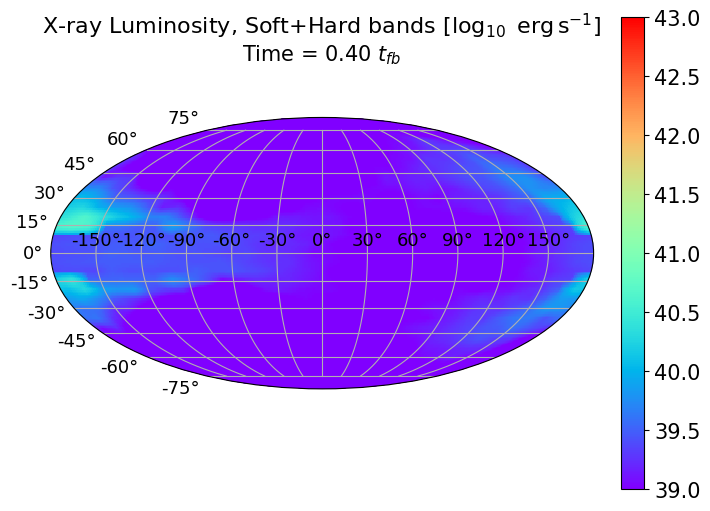}
    \end{minipage}
    \caption{Mollweide sky maps of the Optical+UV luminosity (left) and the Soft+Hard X-ray band luminosities (right), at $t = 0.4\ t_{fb}\approx 1\ \text{day}$.  At this early stage of evolution, very little luminosity is being produced, but hints of later trends have already emerged: optical+UV emission is primarily emitted in the orbital plane where the reprocessing column density is highest, while X-ray emission is suppressed in the orbital plane for the same reason. The pericenter is along the $(0^\circ, 90^\circ)$ ray.}
    \label{fig:skymaps_150}
\end{figure*}

\begin{figure*}
    \begin{minipage}{0.49\textwidth}
    \includegraphics[width=1.0\linewidth]{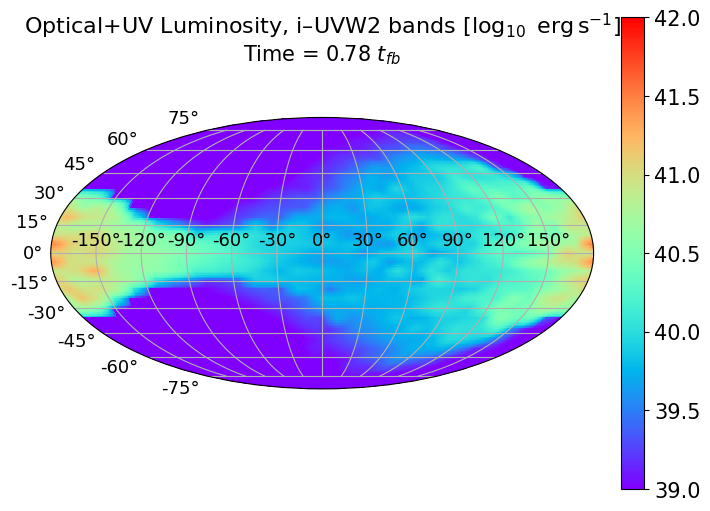} 
    \end{minipage}
    \begin{minipage}{0.49\textwidth}
    \includegraphics[width=1.0\linewidth]{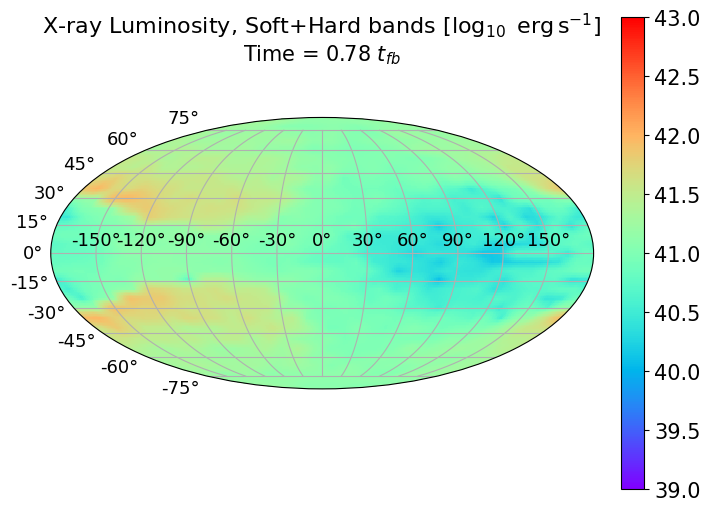}
    \end{minipage}
    \caption{Mollweide sky maps of the Optical+UV luminosity (left) and the Soft+Hard X-ray band luminosities (right), at $t = 0.78\ t_{\rm fb}\approx 2\ \text{days}$, near the peak of the early X-ray flash.  X-ray luminosities have reached near-maximum values, but are suppressed within the orbital plane where they are preferentially absorbed by stellar debris.  Conversely, optical/UV emission is overwhelmingly concentrated in these same planar lines of sight.}
    \label{fig:skymaps_208}
\end{figure*}

\begin{figure*}
    \begin{minipage}{0.49\textwidth}
    \includegraphics[width=1.0\linewidth]{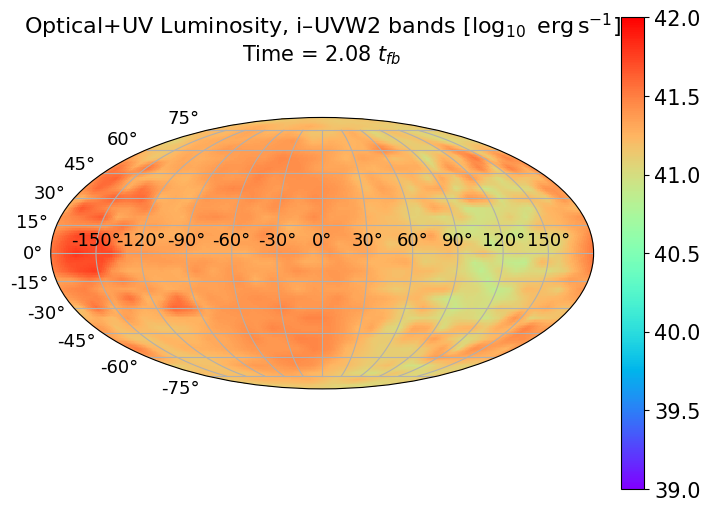} 
    \end{minipage}
    \begin{minipage}{0.49\textwidth}
    \includegraphics[width=1.0\linewidth]{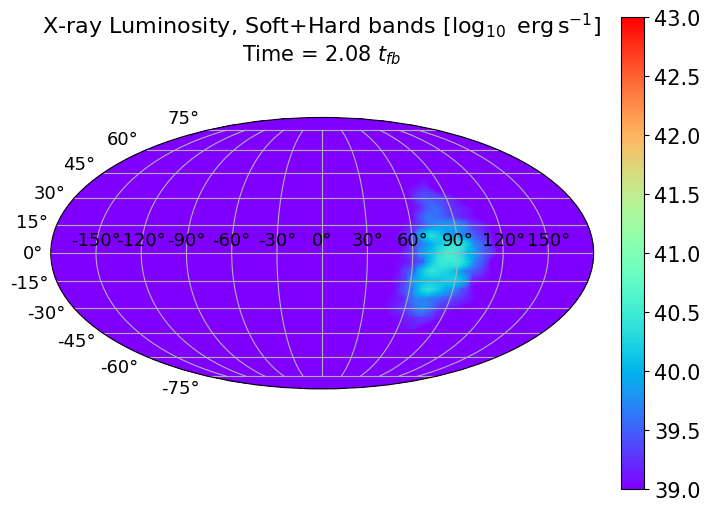}
    \end{minipage}
    \caption{Mollweide sky maps of the Optical+UV luminosity (left) and the Soft+Hard X-ray band luminosities (right), at $t = 2.08\ t_{\rm fb} \approx 5.3\ \text{days}$.  At this late stage, an optically thick photosphere has emerged across $4\pi$ Sr, heavily absorbing X-rays along all lines of sight and producing a more (though not fully) isotropic angular distribution of optical/UV emission.}
    \label{fig:skymaps_400}
\end{figure*}

\begin{figure*}
    \begin{minipage}{0.49\textwidth}
    \includegraphics[width=1.0\textwidth]{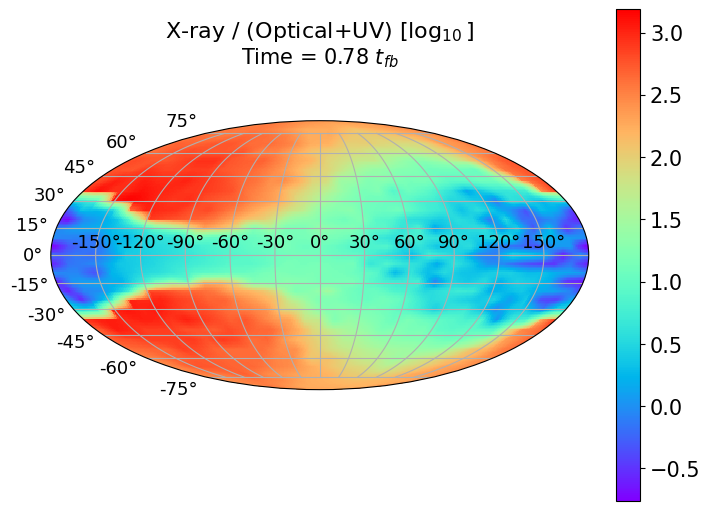} 
    \end{minipage}
    \begin{minipage}{0.49\textwidth}
    \includegraphics[width=1.0\textwidth]{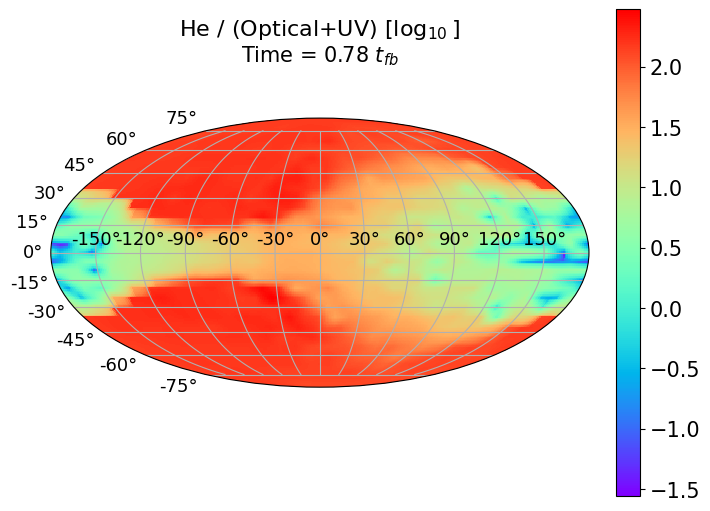}
    \end{minipage}
    \caption{Mollweide sky maps of the ratio between Optical+UV to X-ray luminosity (left) and Optical+UV to our Helium band luminosity (right), at $t=0.79\ t_{fb}\approx 2\ \text{day}$, near peak X-ray luminosity. The optical and UV luminosity is strongest along the orbital plane, where dense material efficiently reprocesses the radiation. In contrast, the X-ray emission originates closer to the black hole and escapes preferentially in directions away from the plane, where the column density is lower and attenuation is reduced.}
    \label{fig:skymaps_ratios}
\end{figure*}

We further define the angle-averaged group luminosity to be 
\begin{equation}    
    L_{\rm g}(t) \equiv \frac{1}{4\pi}\oint_{4\pi} L_{g}(r_{\vec{\Omega}}, \vec{\Omega}, t)d\Omega.
\end{equation}
In Figs.~\ref{fig:skymaps_150}--\ref{fig:skymaps_400} we show isotropic equivalent Mollweide skymaps of the Optical + UV (i-UVW2 bands) and the X-ray (Hard+Soft bands) luminosities at \(t=0.40,\,0.79,\) and \(2.08\,t_{\rm fb}\), respectively. Figure~\ref{fig:skymaps_ratios} also presents ratio maps (Optical+UV/X-ray and Optical+UV/He) at \(t=0.79\,t_{\rm fb}\). Near peak luminosity, the Optical+UV emission is strongest along the disrupted star's orbital plane, consistent with reprocessing of an ionizing continuum in dense debris, whereas the X-ray bands peak at higher inclinations off the plane, where the reprocessing column density is lower (see especially Figs.~\ref{fig:skymaps_208},~\ref{fig:skymaps_ratios}). At \(t=2.08\,t_{\rm fb}\) a localized X-ray bright spot appears (Fig.~\ref{fig:skymaps_400}), density and temperature slices indicate it arises from hot gas breaking out of the inner disk (Figs.~\ref{fig:TDE_Xray_density_400}--\ref{fig:TDE_temperatue_400}). 

\begin{figure*}
    \centering
    \includegraphics[width=0.75\textwidth]{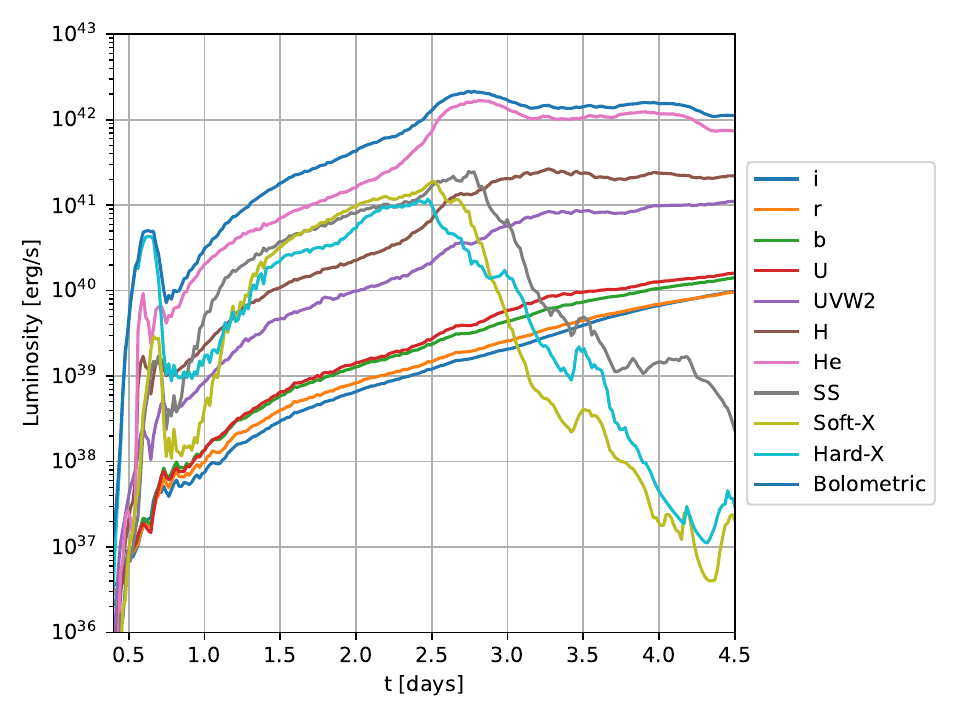}
    \caption{Light curves for the different energy groups.  Most radiation emerges in UV wavelengths; primarily in the unobservable EUV bands (H and He).  UV bands (including {\it Swift} UVW2) reach peaks or plateaus near $t\approx 1 t_{\rm fb}$, while optical wavelengths monotonically rise for the duration of the simulation.  The earliest X-ray flash (at $t\approx 0.6 ~{\rm day}$) is likely an unphysical artifact associated with under-resolving the stream tip due to its low density, as it returns to pericenter at early times \citep{Martire+25}, but the broader X-ray flare from $\approx 1$ to $\approx 2.5$ days is more likely physical, and agrees qualitatively with both the simulation of \cite{steinberg_streamdisk_2024} and observations of the TDE AT 2022dsb \citep{malyali_transient_2024}.  This paper's multigroup simulations are the first self-consistent prediction of an early X-ray flash in an end-to-end simulation, suggesting that these flares may offer observational insight into both circularization mechanisms and viewing angles.}
    \label{fig:TDE_lightcurves}
\end{figure*}

\begin{figure}
    \centering
    \subfloat[a][]{
        \includegraphics[width=0.95\linewidth]{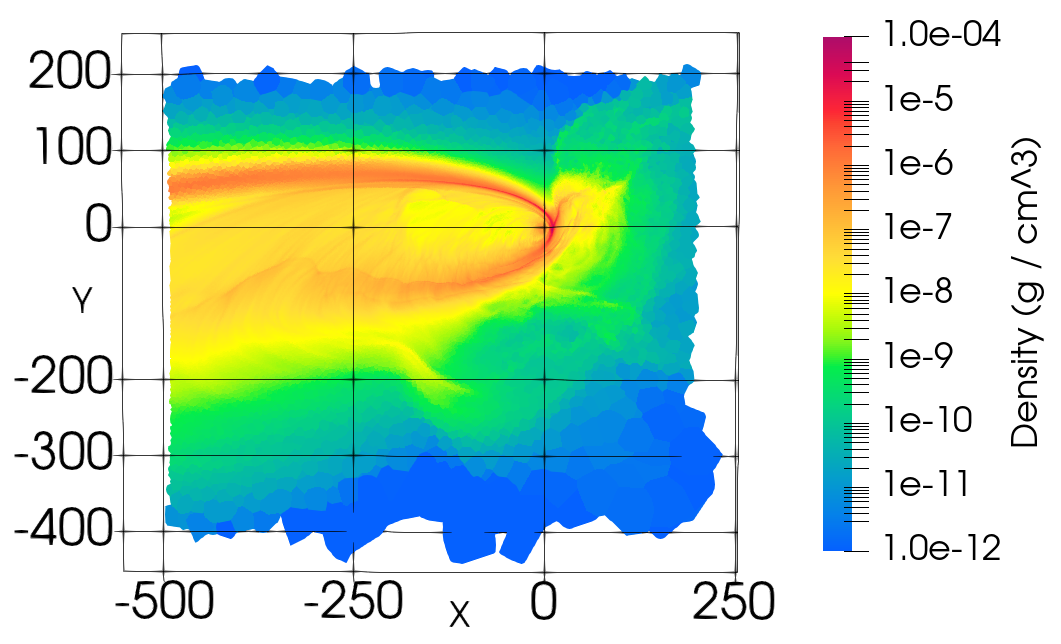}
         \label{fig:TDE_density_400}
    } \\
    \subfloat[b][]{
        \includegraphics[width=0.95\linewidth]{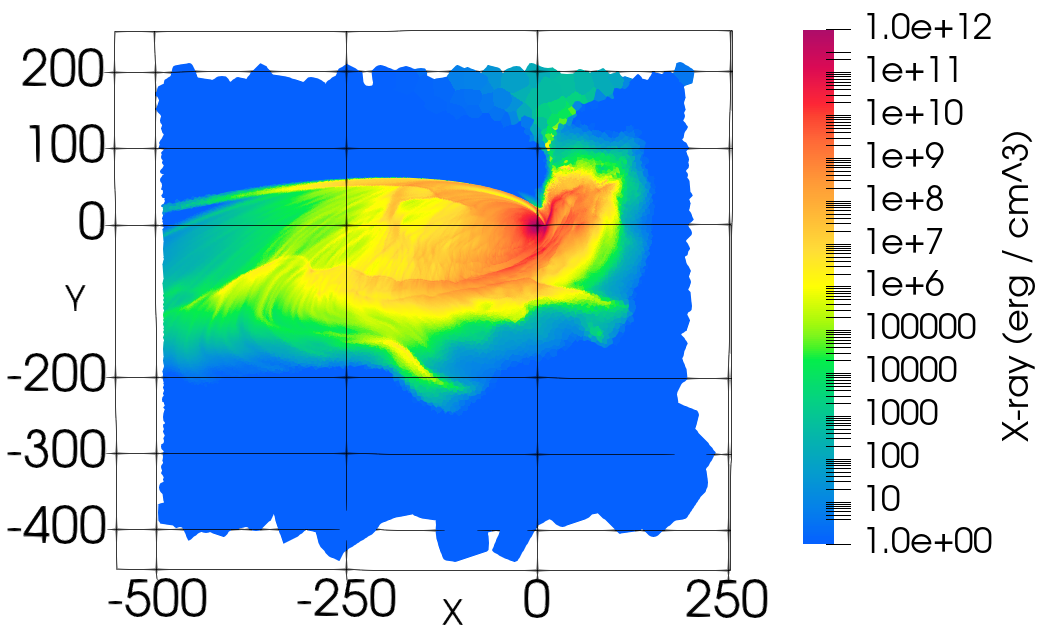}
         \label{fig:TDE_Xray_400}
    }
    \caption{(a) Density slice in the orbital plane at $t=2.08 t_{\rm fb} (\approx 5.3\ \text{day})$ after the star was disrupted. (b) Slice in the orbital plane of the X-ray energy density, taken at  the same time. 
    The pericentric nozzle shock has generated a complex and turbulent flow, even though little circularization has occurred in absolute terms.  A thin, high-temperature band parallels a density jump extending upwards (along the in-plane $y$-axis) from the nozzle shock.       
    }
    \label{fig:TDE_Xray_density_400}
\end{figure}

\begin{figure*}
    \centering
    \includegraphics[width=0.75\textwidth]{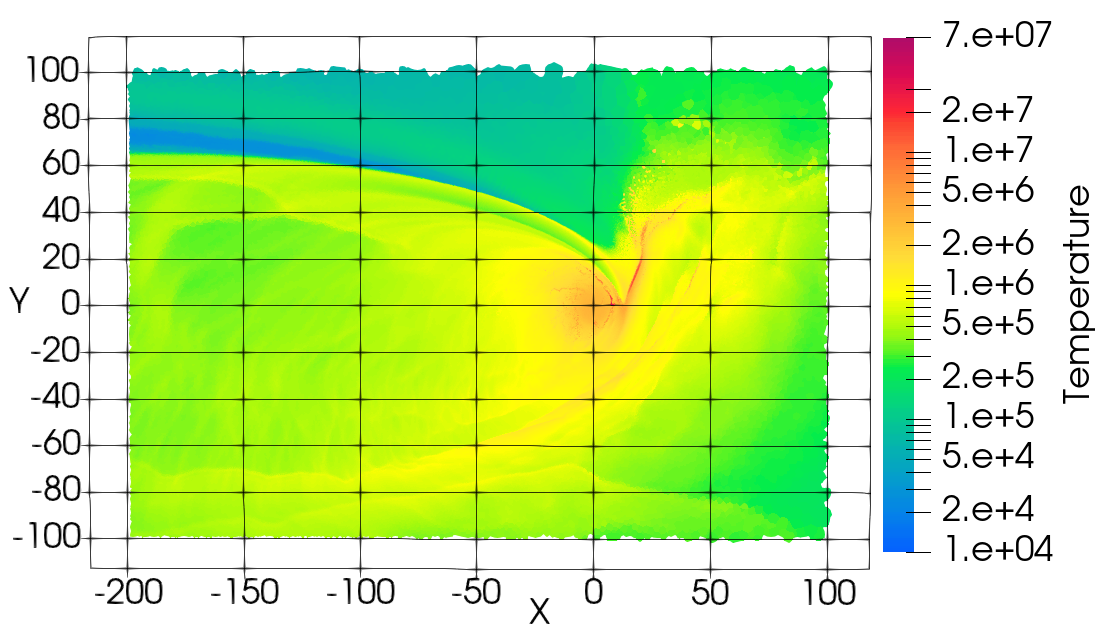}
    \caption{Gas temperature slice in the orbital plane at  $t=2.08\ t_{fb} (\approx 5.3\ \text{days})$, zoomed on the pericenter and  ``breakout'' region.  The stream of hot gas visible on the positive $y$-axis is the origin of the residual late-time X-ray emission visible in Fig. \ref{fig:skymaps_400}, illustrating the complex viewing angle dependence of high-energy radiation at early times in TDEs.}
    \label{fig:TDE_temperatue_400}
\end{figure*}

We further quantify the observable features of this simulated TDE by plotting sky-averaged multiband light curves, in Fig. \ref{fig:TDE_lightcurves}.  The sky-averaged light curves exhibit a clear X‑ray flash, peaking at $\approx2.5$ days after disruption (Fig. \ref{fig:TDE_lightcurves}). This result appears consistent with earlier predictions of similar X-ray flashes in supermassive black hole TDEs from \cite{steinberg_streamdisk_2024}, which appear to have been observed in the (supermassive) TDE AT 2022dsb \citep{malyali_transient_2024}. We note the presence of a brief, earlier spike at $t\approx 0.6\ \text{days}$.  As stated in the caption of Fig.~\ref{fig:TDE_lightcurves}, we regard this earliest spike as likely being a numerical artifact associated with under-resolution of the returning stream tip near pericenter at early times, when its density is low.  Similar early flashes of radiation from the stream tip were seen in preliminary, under-resolved runs in \citet{Martire+25}. In contrast to \citet{steinberg_streamdisk_2024}, we have made self-consistent predictions for the time-dependent X-ray luminosity in this paper using our novel multigroup FLD transfer module; with grey simulations, spectral predictions can only be made in post-processing, which may not be self-consistent in situations like TDE circularization where (frequency-dependent) radiation pressure dominates over purely hydrodynamical forces.

Both this multigroup simulation and the post-processing of grey RHD results in \citet{steinberg_streamdisk_2024} suggest that early X-ray flashes can be produced by the activity of the nozzle shock at times (and viewing angles) when the column density of colder absorbing debris is not yet too large.  This X-ray flash merits further investigation: if detected in future samples of TDEs, it could serve as a ``starting gun,'' timing the onset of mass return to pericenter.  As many of the thorniest theoretical questions concerning TDEs are downstream from long-standing uncertainties concerning the duration of the circularization process, a concrete measurement of the time when circularization begins would be highly valuable.

In agreement with some prior theoretical work \citep{LuKumar18, Dai+18, Wen+20}, the bulk of our flare's bolometric luminosity emerges at unobservable EUV wavelengths (i.e. our H and He frequency groups, to which the Milky Way's ISM is optically thick in absorption).  These frequency groups track the X-ray band evolution at early times (up to $\approx 1 t_{\rm fb}$, and peak at timescales comparable to the early X-ray peak, but they do not suffer the same rapid, multiple order of magnitude drop seen in X-rays, likely due to intermediate temperature gas (i.e. gas that thermally emits in the EUV) at the color surface.  At even longer wavelengths, Fig. \ref{fig:TDE_lightcurves} shows a monotonic increase in optical and near-UV luminosity as a function of time.

The dissipation dynamics in this simulation are complex, which we illustrate in Fig. \ref{fig:TDE_Xray_density_400} using midplane density and temperature slices taken towards the end of the simulation.  Visual inspection of the density map shows that even though little circularization has occurred (in agreement with the grey RHD IMBH simulation of \citealt{Martire+25}) due to the general weakness of the nozzle shock \citep{Guillochon+14, Hu+25, Andalman+25}, the flow has become turbulent.  Gas temperatures are highest in the central regions of the eccentric accretion flow, although a thin band or stream of high-temperature gas just upwards (in-plane, along the positive $y$-axis), as can be better seen in the Fig. \ref{fig:TDE_temperatue_400} zoom-in.  This hot stream is ultimately responsible for the limited remaining late-time X-ray emission, as can be seen by comparing to the Mollweide plot in Fig. \ref{fig:skymaps_400}.  Early-time emission from TDEs likely exhibits a complex viewing angle dependence, especially for harder photon energies most susceptible to absorption and reprocessing.

A deeper investigation of multigroup simulations of TDEs will be presented in Giron et al {\it in prep}, but the pilot simulation shown here illustrates the potential of our new module for making directly testable predictions about TDE emission.  While this paper was in final stages of preparation, \citealt{Huang+25b} published a different multigroup RHD simulation of a TDE.  While these two papers are substantially different from each other (in contrast to our end-to-end IMBH TDE simulation, \citealt{Huang+25b} present a SMBH TDE simulation that skips over stream dynamics by using a mass injection scheme similar to that in \citealt{Huang+25}), the conclusions about the angular dependence of different frequency groups appear qualitatively similar.  Notably, both works confirm the earlier prediction of \citet{steinberg_streamdisk_2024} regarding early X-ray emission originating from shock dissipation.

\section{Conclusions} \label{sec:conclusions}
We have modified the moving-mesh radiation hydrodynamics code \textsc{rich} to incorporate a multigroup module for radiation transfer.  This builds on our previous work, adding grey mixed frame FLD radiation transfer into \textsc{rich}, the results of which were previously presented in \citet{steinberg_streamdisk_2024}. The updated version of \textsc{rich} is publicly available, in keeping with our longstanding policy of accessible code development.  \textsc{rich} is now the only moving-mesh RHD code capable of transporting radiation using multigroup photons, 
creating a unique capability for astrophysical problems where (i) radiation is dynamically important and (ii) the relevant radiation field spans orders of magnitude in photon energy.  The updated \textsc{rich} will also be valuable for general problems in astrophysical hydrodynamics where broadband spectral predictions are useful, reducing the need for post-processing of hydrodynamic simulations with dedicated multigroup radiation transfer codes.

Our implementation of multigroup FLD into \textsc{rich} expands the capabilities of moving-mesh RHD at the cost of greater complexity and computational expense.  The multi-step RHD algorithm is described in more detail in \S \ref{sec:implementation}, but in brief, the code first updates the hydrodynamic variables, then moves the Voronoi mesh to follow the bulk fluid flow, and then updates the FLD radiation equation\footnote{This overview is slightly oversimplified; for example, radiation advection occurs during the hydrodynamics sub-step.}.  Finally, radiation forces update the gas momentum and energy variables.  To mitigate the greater computational cost of performing the RHD in a multigroup way, we have implemented a novel scheme to accelerate convergence in optically thick cells, with minimal loss of accuracy, by limiting the absorption coefficients (as can be seen in Fig. \ref{fig:convergence}).   

We have validated both the grey and the multigroup RHD modules in \textsc{rich} in a series of test problems, presented in \S \ref{sec:grey_validation} and \S \ref{sec:multigroup_validation}, respectively.  In general, the numerical solutions from \textsc{rich} achieve excellent agreement with previously known self-similar test problems and analytic jump conditions for radiative shocks, as well as a zero-dimensional test problem we have constructed to probe the Doppler term in multigroup RHD.

As a final test for multigroup \textsc{rich}, we have gone beyond test problems with analytic solutions and have run the first end-to-end three dimensional simulation of a TDE in multigroup RHD.  Our simulation follows the disruption of a lower main sequence star by a $M=10^4 M_\odot$ IMBH, using astrophysical parameters identical to the grey RHD simulations recently presented in \citet{Martire+25}.  This simulation produced multi-band light curves across 10 different frequency groups, and runs for a little over 2 fallback times.  Although we will defer more detailed explorations of multigroup TDE simulations for future work (Giron et al. {\it in prep}), this proof-of-principle simulation is already quite interesting.  The bulk hydrodynamics and bolometric light curve appear broadly consistent with the slow circularization seen in \citet{Martire+25}, but the emergent radiation field exhibits a complex angular dependence and frequency evolution.  Optical and near-UV bands rise monotonically but never make up more than $\sim 1\%$ of the bolometric emission, which is released primarily in the unobservable EUV (although we note that the far-UV {\it Swift} UVW2 band does eventually reach $\sim 10\%$ of the bolometric luminosity).  

The light curve evolution at short wavelengths is richer: both soft and hard X-ray bands rise rapidly in tandem with bolometric luminosity, and at the peak of the bolometric light curve (achieved around $t\approx 1 t_{\rm fb}$), a few tens of percent of the emitted radiation is in X-rays.  However, X-rays can only escape along lines of sight with low gas column density, and the partial isotropization of the stellar debris quickly closes the window on X-ray emission.  The X-ray light curve peaks near $1 t_{\rm fb}$ and then enters a rapid, multiple orders of magnitude decline, although we anticipate it will eventually rise again due to the onset of central accretion (not treated in this simulation).  This early time, pre-peak X-ray flash qualitatively agrees with the SMBH TDE predictions of \citet{steinberg_streamdisk_2024}, although here we have produced our spectral predictions in a more self-consistent way.  This early X-ray flash, which ultimately originates from shock dissipation in the first stages of the circularization process, may already have been seen in one TDE\footnote{Immediately prior to completion of this paper, an analysis of the {\it eROSITA} X-ray selected TDE sample appears to have uncovered a second TDE with an X-ray flash prior to optical peak,  SRGe J131014.7+444319 \citep{Zhang+25}.} \citep{malyali_transient_2024}, and could provide direct observational constraints on the circularization efficiency if it could be studied in more detail in a broader TDE sample.

We note two limitations of the present comoving-frame FLD formulation. 
Because a conserved total (gas+radiation) energy is defined only in an inertial frame, the comoving-frame equations do not admit a manifestly conservative form for the combined energy budget. This is especially undesirable for dynamically evolving meshes such as the moving Voronoi grid used by \textsc{rich} 
(cf.\ \citealt{Krumholz+07b}). In addition, we truncate the system at first order in $v/c$, neglecting $\mathcal{O}(v^2/c^2)$ (and higher) corrections that can become leading-order in the dynamic-diffusion regime and in nonuniform, time-dependent flows 
\citep[\S\,3.1]{Krumholz+07b, Krumholz_2024}.

In the future, we hope that a publicly available, multigroup moving-mesh RHD code may provide broad utility in astrophysics.  We anticipate that the radiation transfer module in \textsc{rich} will be particularly useful for simulating astrophysical problems where (i) radiation forces become dynamically important and (ii) some combination of large dynamic range and highly supersonic flow pose severe challenges for more standard numerical hydrodynamics schemes.  Multigroup \textsc{rich} will be especially applicable for the subset of these problems where broad-band light curves or spectral predictions are desired.  Within our own collaboration, we anticipate extending the pilot multigroup TDE simulation presented here to a broader range of TDE parameters.  In the very near future, both the Vera Rubin Observatory \citep{BricmanGomboc20} and the Israeli-NASA-DESY {\it ULTRASAT} UV survey \citep{Shvartzvald+24} will likely discover thousands of TDEs, but the analysis of these flares could be gravely limited by the lack of first-principles models.  A broader parameter survey of multigroup RHD TDE simulations may be a necessary step in realizing the scientific potential of TDEs in this decade.

\section*{Acknowledgement}
We thank P. Chang, P. Martire, and E.M. Rossi for fruitful discussions. NCS acknowledges support from the Binational Science Foundation (grant No. 2020397) and the Israel Science Foundation (Individual Research Grant No. 2414/23). The simulations were performed on the ICPL cluster at the Racah Institute of Physics as well as the Dutch national e-infrastructure with the support of the SURF Cooperative using grant no. EINF-13675 and NWO-2022.016.

\bibliography{bib}{}

@ARTICLE{Dai+18,
       author = {{Dai}, Lixin and {McKinney}, Jonathan C. and {Roth}, Nathaniel and {Ramirez-Ruiz}, Enrico and {Miller}, M. Coleman},
        title = "{A Unified Model for Tidal Disruption Events}",
      journal = {\apjl},
         year = 2018,
        month = jun,
       volume = {859},
       number = {2},
          eid = {L20},
        pages = {L20},
          doi = {10.3847/2041-8213/aab429},
archivePrefix = {arXiv},
       eprint = {1803.03265},
 primaryClass = {astro-ph.HE},
       adsurl = {https://ui.adsabs.harvard.edu/abs/2018ApJ...859L..20D}
}

@ARTICLE{Wen+20,
       author = {{Wen}, Sixiang and {Jonker}, Peter G. and {Stone}, Nicholas C. and {Zabludoff}, Ann I. and {Psaltis}, Dimitrios},
        title = "{Continuum-fitting the X-Ray Spectra of Tidal Disruption Events}",
      journal = {\apj},
         year = 2020,
        month = jul,
       volume = {897},
       number = {1},
          eid = {80},
        pages = {80},
          doi = {10.3847/1538-4357/ab9817},
archivePrefix = {arXiv},
       eprint = {2003.12583},
 primaryClass = {astro-ph.HE},
       adsurl = {https://ui.adsabs.harvard.edu/abs/2020ApJ...897...80W}
}

@ARTICLE{LuKumar18,
       author = {{Lu}, Wenbin and {Kumar}, Pawan},
        title = "{On the Missing Energy Puzzle of Tidal Disruption Events}",
      journal = {\apj},
         year = 2018,
        month = oct,
       volume = {865},
       number = {2},
          eid = {128},
        pages = {128},
          doi = {10.3847/1538-4357/aad54a},
archivePrefix = {arXiv},
       eprint = {1802.02151},
 primaryClass = {astro-ph.HE},
       adsurl = {https://ui.adsabs.harvard.edu/abs/2018ApJ...865..128L}
}

@ARTICLE{Huang+24,
       author = {{Huang}, Xiaoshan and {Davis}, Shane W. and {Jiang}, Yan-fei},
        title = "{Pre-peak Emission in Tidal Disruption Events}",
      journal = {\apj},
         year = 2024,
        month = oct,
       volume = {974},
       number = {2},
          eid = {165},
        pages = {165},
          doi = {10.3847/1538-4357/ad6c39},
archivePrefix = {arXiv},
       eprint = {2404.18446},
 primaryClass = {astro-ph.HE},
       adsurl = {https://ui.adsabs.harvard.edu/abs/2024ApJ...974..165H}
}

@ARTICLE{Andalman+22,
       author = {{Andalman}, Zachary L. and {Liska}, Matthew T.~P. and {Tchekhovskoy}, Alexander and {Coughlin}, Eric R. and {Stone}, Nicholas},
        title = "{Tidal disruption discs formed and fed by stream-stream and stream-disc interactions in global GRHD simulations}",
      journal = {\mnras},
         year = 2022,
        month = feb,
       volume = {510},
       number = {2},
        pages = {1627-1648},
          doi = {10.1093/mnras/stab3444},
archivePrefix = {arXiv},
       eprint = {2008.04922},
 primaryClass = {astro-ph.HE},
       adsurl = {https://ui.adsabs.harvard.edu/abs/2022MNRAS.510.1627A}
}

@ARTICLE{Hayasaki+13,
       author = {{Hayasaki}, Kimitake and {Stone}, Nicholas and {Loeb}, Abraham},
        title = "{Finite, intense accretion bursts from tidal disruption of stars on bound orbits}",
      journal = {\mnras},
         year = 2013,
        month = sep,
       volume = {434},
       number = {2},
        pages = {909-924},
          doi = {10.1093/mnras/stt871},
archivePrefix = {arXiv},
       eprint = {1210.1333},
 primaryClass = {astro-ph.HE},
       adsurl = {https://ui.adsabs.harvard.edu/abs/2013MNRAS.434..909H}
}

@ARTICLE{Piran+15,
       author = {{Piran}, Tsvi and {Svirski}, Gilad and {Krolik}, Julian and {Cheng}, Roseanne M. and {Shiokawa}, Hotaka},
        title = "{‧Disk Formation Versus Disk Accretion{\textemdash}What Powers Tidal Disruption Events?}",
      journal = {\apj},
         year = 2015,
        month = jun,
       volume = {806},
       number = {2},
          eid = {164},
        pages = {164},
          doi = {10.1088/0004-637X/806/2/164},
archivePrefix = {arXiv},
       eprint = {1502.05792},
 primaryClass = {astro-ph.HE},
       adsurl = {https://ui.adsabs.harvard.edu/abs/2015ApJ...806..164P}
}

@ARTICLE{Shiokawa+15,
       author = {{Shiokawa}, Hotaka and {Krolik}, Julian H. and {Cheng}, Roseanne M. and {Piran}, Tsvi and {Noble}, Scott C.},
        title = "{General Relativistic Hydrodynamic Simulation of Accretion Flow from a Stellar Tidal Disruption}",
      journal = {\apj},
         year = 2015,
        month = may,
       volume = {804},
       number = {2},
          eid = {85},
        pages = {85},
          doi = {10.1088/0004-637X/804/2/85},
archivePrefix = {arXiv},
       eprint = {1501.04365},
 primaryClass = {astro-ph.HE},
       adsurl = {https://ui.adsabs.harvard.edu/abs/2015ApJ...804...85S}
}

@ARTICLE{BonnerotStone21,
       author = {{Bonnerot}, C. and {Stone}, N.~C.},
        title = "{Formation of an Accretion Flow}",
      journal = {\ssr},
         year = 2021,
        month = feb,
       volume = {217},
       number = {1},
          eid = {16},
        pages = {16},
          doi = {10.1007/s11214-020-00789-1},
archivePrefix = {arXiv},
       eprint = {2008.11731},
 primaryClass = {astro-ph.HE},
       adsurl = {https://ui.adsabs.harvard.edu/abs/2021SSRv..217...16B}
}

@ARTICLE{vanVelzen+20,
       author = {{van Velzen}, Sjoert and {Holoien}, Thomas W.-S. and {Onori}, Francesca and {Hung}, Tiara and {Arcavi}, Iair},
        title = "{Optical-Ultraviolet Tidal Disruption Events}",
      journal = {\ssr},
         year = 2020,
        month = oct,
       volume = {216},
       number = {8},
          eid = {124},
        pages = {124},
          doi = {10.1007/s11214-020-00753-z},
archivePrefix = {arXiv},
       eprint = {2008.05461},
 primaryClass = {astro-ph.HE},
       adsurl = {https://ui.adsabs.harvard.edu/abs/2020SSRv..216..124V}
}

@ARTICLE{Krolik+20,
       author = {{Krolik}, Julian H. and {Armitage}, Philip J. and {Jiang}, Yanfei and {Lodato}, Giuseppe},
        title = "{Future Simulations of Tidal Disruption Events}",
      journal = {\ssr},
         year = 2020,
        month = jul,
       volume = {216},
       number = {5},
          eid = {88},
        pages = {88},
          doi = {10.1007/s11214-020-00680-z},
archivePrefix = {arXiv},
       eprint = {2006.03693},
 primaryClass = {astro-ph.HE},
       adsurl = {https://ui.adsabs.harvard.edu/abs/2020SSRv..216...88K}
}

@ARTICLE{BricmanGomboc20,
       author = {{Bricman}, Katja and {Gomboc}, Andreja},
        title = "{The Prospects of Observing Tidal Disruption Events with the Large Synoptic Survey Telescope}",
      journal = {\apj},
         year = 2020,
        month = feb,
       volume = {890},
       number = {1},
          eid = {73},
        pages = {73},
          doi = {10.3847/1538-4357/ab6989},
archivePrefix = {arXiv},
       eprint = {1906.08235},
 primaryClass = {astro-ph.HE},
       adsurl = {https://ui.adsabs.harvard.edu/abs/2020ApJ...890...73B}
}

@ARTICLE{Shvartzvald+24,
       author = {{Shvartzvald}, Y. and {Waxman}, E. and {Gal-Yam}, A. and {Ofek}, E.~O. and {Ben-Ami}, S. and {Berge}, D. and {Kowalski}, M. and {B{\"u}hler}, R. and {Worm}, S. and {Rhoads}, J.~E. and {Arcavi}, I. and {Maoz}, D. and {Polishook}, D. and {Stone}, N. and {Trakhtenbrot}, B. and {Ackermann}, M. and {Aharonson}, O. and {Birnholtz}, O. and {Chelouche}, D. and {Guetta}, D. and {Hallakoun}, N. and {Horesh}, A. and {Kushnir}, D. and {Mazeh}, T. and {Nordin}, J. and {Ofir}, A. and {Ohm}, S. and {Parsons}, D. and {Pe'er}, A. and {Perets}, H.~B. and {Perdelwitz}, V. and {Poznanski}, D. and {Sadeh}, I. and {Sagiv}, I. and {Shahaf}, S. and {Soumagnac}, M. and {Tal-Or}, L. and {Santen}, J. Van and {Zackay}, B. and {Guttman}, O. and {Rekhi}, P. and {Townsend}, A. and {Weinstein}, A. and {Wold}, I.},
        title = "{ULTRASAT: A Wide-field Time-domain UV Space Telescope}",
      journal = {\apj},
         year = 2024,
        month = mar,
       volume = {964},
       number = {1},
          eid = {74},
        pages = {74},
          doi = {10.3847/1538-4357/ad2704},
archivePrefix = {arXiv},
       eprint = {2304.14482},
 primaryClass = {astro-ph.IM},
       adsurl = {https://ui.adsabs.harvard.edu/abs/2024ApJ...964...74S}
}

@ARTICLE{Zhang+25,
       author = {{Zhang}, Zirui and {Yao}, Yuhan and {Gilfanov}, Marat and {Sazonov}, Sergey and {Medvedev}, Pavel and {Khorunzhev}, Georgii and {Sunyaev}, Rashid and {Ravi}, Vikram and {Kulkarni}, S.~R. and {Somalwar}, Jean and {Chornock}, Ryan and {Bikmaev}, Ilfan and {Gorbachev}, Mark A.},
        title = "{eROSITA-RU Tidal Disruption Events with Keck-I/LRIS: Sample Selection, Optical Properties, and Host Galaxy Demographics}",
      journal = {arXiv e-prints},
         year = 2025,
        month = dec,
          eid = {arXiv:2512.12480},
        pages = {arXiv:2512.12480},
archivePrefix = {arXiv},
       eprint = {2512.12480},
 primaryClass = {astro-ph.HE},
       adsurl = {https://ui.adsabs.harvard.edu/abs/2025arXiv251212480Z}
}

@ARTICLE{Hu+25,
       author = {{Fitz Hu}, Fangyi and {Mandel}, Ilya and {Nealon}, Rebecca and {Price}, Daniel J.},
        title = "{Converged simulations of the nozzle shock in tidal disruption events}",
      journal = {arXiv e-prints},
         year = 2025,
        month = oct,
          eid = {arXiv:2510.04790},
        pages = {arXiv:2510.04790},
          doi = {10.48550/arXiv.2510.04790},
archivePrefix = {arXiv},
       eprint = {2510.04790},
 primaryClass = {astro-ph.HE},
       adsurl = {https://ui.adsabs.harvard.edu/abs/2025arXiv251004790F}
}

@ARTICLE{Guillochon+14,
       author = {{Guillochon}, James and {Manukian}, Haik and {Ramirez-Ruiz}, Enrico},
        title = "{PS1-10jh: The Disruption of a Main-sequence Star of Near-solar Composition}",
      journal = {\apj},
         year = 2014,
        month = mar,
       volume = {783},
       number = {1},
          eid = {23},
        pages = {23},
          doi = {10.1088/0004-637X/783/1/23},
archivePrefix = {arXiv},
       eprint = {1304.6397},
 primaryClass = {astro-ph.HE},
       adsurl = {https://ui.adsabs.harvard.edu/abs/2014ApJ...783...23G}
}

@ARTICLE{Andalman+25,
       author = {{Andalman}, Zachary L. and {Quataert}, Eliot and {Coughlin}, Eric R. and {Nixon}, C.~J.},
        title = "{Resolving the (Debate About) Nozzle Shocks in Tidal Disruption Events}",
      journal = {arXiv e-prints},
         year = 2025,
        month = dec,
          eid = {arXiv:2512.08928},
        pages = {arXiv:2512.08928},
          doi = {10.48550/arXiv.2512.08928},
archivePrefix = {arXiv},
       eprint = {2512.08928},
 primaryClass = {astro-ph.HE},
       adsurl = {https://ui.adsabs.harvard.edu/abs/2025arXiv251208928A}
}

@ARTICLE{Huang+25b,
       author = {{Huang}, Xiaoshan and {Meza}, Maria Renee and {Yun}, Sol Bin and {Mockler}, Brenna and {Davis}, Shane W. and {Jiang}, Yan-fei},
        title = "{X-ray Variability and Photosphere Evolution during Accretion Disk Formation in Tidal Disruption Events}",
      journal = {arXiv e-prints},
         year = 2025,
        month = dec,
          eid = {arXiv:2512.12985},
        pages = {arXiv:2512.12985},
archivePrefix = {arXiv},
       eprint = {2512.12985},
 primaryClass = {astro-ph.HE},
       adsurl = {https://ui.adsabs.harvard.edu/abs/2025arXiv251212985H}
}

@ARTICLE{Martire+25,
       author = {{Martire}, Paola and {Rossi}, Elena Maria and {Chamberlain Stone}, Nicholas and {Steinberg}, Elad and {Kilmetis}, Konstantinos and {Linial}, Itai},
        title = "{Wind-mediated Eddington-limited emission in a 1e4 Black Hole Tidal Disruption Event}",
      journal = {arXiv e-prints},
         year = 2025,
        month = dec,
          eid = {arXiv:2512.10564},
        pages = {arXiv:2512.10564},
          doi = {10.48550/arXiv.2512.10564},
archivePrefix = {arXiv},
       eprint = {2512.10564},
 primaryClass = {astro-ph.HE},
       adsurl = {https://ui.adsabs.harvard.edu/abs/2025arXiv251210564M}
}

@ARTICLE{Kannan+19,
       author = {{Kannan}, Rahul and {Vogelsberger}, Mark and {Marinacci}, Federico and {McKinnon}, Ryan and {Pakmor}, R{\"u}diger and {Springel}, Volker},
        title = "{AREPO-RT: radiation hydrodynamics on a moving mesh}",
      journal = {\mnras},
         year = 2019,
        month = may,
       volume = {485},
       number = {1},
        pages = {117-149},
          doi = {10.1093/mnras/stz287},
archivePrefix = {arXiv},
       eprint = {1804.01987},
 primaryClass = {astro-ph.IM},
       adsurl = {https://ui.adsabs.harvard.edu/abs/2019MNRAS.485..117K}
}

@ARTICLE{Ma+25,
       author = {{Ma}, Jing-Ze and {Pakmor}, R\textbackslash''udiger and {Justham}, Stephen and {de Mink}, Selma E.},
        title = "{AREPO-IDORT: Implicit Discrete Ordinates Radiation Transport for Radiation Magnetohydrodynamics on an Unstructured Moving Mesh}",
      journal = {arXiv e-prints},
         year = 2025,
        month = mar,
          eid = {arXiv:2503.16627},
        pages = {arXiv:2503.16627},
          doi = {10.48550/arXiv.2503.16627},
archivePrefix = {arXiv},
       eprint = {2503.16627},
 primaryClass = {astro-ph.IM},
       adsurl = {https://ui.adsabs.harvard.edu/abs/2025arXiv250316627M}
}

@ARTICLE{Chang+20,
       author = {{Chang}, Philip and {Davis}, Shane W. and {Jiang}, Yan-Fei},
        title = "{Time-dependent radiation hydrodynamics on a moving mesh}",
      journal = {\mnras},
         year = 2020,
        month = apr,
       volume = {493},
       number = {4},
        pages = {5397-5407},
          doi = {10.1093/mnras/staa573},
archivePrefix = {arXiv},
       eprint = {2002.08377},
 primaryClass = {astro-ph.IM},
       adsurl = {https://ui.adsabs.harvard.edu/abs/2020MNRAS.493.5397C}
}

@ARTICLE{SteinbergMetzger18,
       author = {{Steinberg}, Elad and {Metzger}, Brian D.},
        title = "{The multidimensional structure of radiative shocks: suppressed thermal X-rays and relativistic ion acceleration}",
      journal = {\mnras},
         year = 2018,
        month = sep,
       volume = {479},
       number = {1},
        pages = {687-702},
          doi = {10.1093/mnras/sty1641},
archivePrefix = {arXiv},
       eprint = {1805.03223},
 primaryClass = {astro-ph.HE},
       adsurl = {https://ui.adsabs.harvard.edu/abs/2018MNRAS.479..687S}
}

@ARTICLE{Nelson+13,
       author = {{Nelson}, Dylan and {Vogelsberger}, Mark and {Genel}, Shy and {Sijacki}, Debora and {Kere{\v{s}}}, Du{\v{s}}an and {Springel}, Volker and {Hernquist}, Lars},
        title = "{Moving mesh cosmology: tracing cosmological gas accretion}",
      journal = {\mnras},
         year = 2013,
        month = mar,
       volume = {429},
       number = {4},
        pages = {3353-3370},
          doi = {10.1093/mnras/sts595},
archivePrefix = {arXiv},
       eprint = {1301.6753},
 primaryClass = {astro-ph.CO},
       adsurl = {https://ui.adsabs.harvard.edu/abs/2013MNRAS.429.3353N}
}

@ARTICLE{ZierSpringel22,
       author = {{Zier}, Oliver and {Springel}, Volker},
        title = "{Simulating cold shear flows on a moving mesh}",
      journal = {\mnras},
         year = 2022,
        month = sep,
       volume = {515},
       number = {1},
        pages = {525-542},
          doi = {10.1093/mnras/stac1783},
archivePrefix = {arXiv},
       eprint = {2205.07916},
 primaryClass = {astro-ph.IM},
       adsurl = {https://ui.adsabs.harvard.edu/abs/2022MNRAS.515..525Z}
}

@ARTICLE{Steinberg+16,
       author = {{Steinberg}, Elad and {Yalinewich}, Almog and {Sari}, Re'em},
        title = "{Grid noise in moving mesh codes: fixing the volume inconsistency problem}",
      journal = {\mnras},
         year = 2016,
        month = jun,
       volume = {459},
       number = {2},
        pages = {1596-1601},
          doi = {10.1093/mnras/stw783},
archivePrefix = {arXiv},
       eprint = {1603.06753},
 primaryClass = {physics.comp-ph},
       adsurl = {https://ui.adsabs.harvard.edu/abs/2016MNRAS.459.1596S}
}

@ARTICLE{Steinberg+15,
       author = {{Steinberg}, Elad and {Yalinewich}, Almog and {Sari}, Re'em and {Duffell}, Paul},
        title = "{Balancing the Load: A Voronoi Based Scheme for Parallel Computations}",
      journal = {\apjs},
         year = 2015,
        month = jan,
       volume = {216},
       number = {1},
          eid = {14},
        pages = {14},
          doi = {10.1088/0067-0049/216/1/14},
archivePrefix = {arXiv},
       eprint = {1408.3196},
 primaryClass = {astro-ph.IM},
       adsurl = {https://ui.adsabs.harvard.edu/abs/2015ApJS..216...14S}
}

@ARTICLE{AubertTeyssier08,
       author = {{Aubert}, Dominique and {Teyssier}, Romain},
        title = "{A radiative transfer scheme for cosmological reionization based on a local Eddington tensor}",
      journal = {\mnras},
         year = 2008,
        month = jun,
       volume = {387},
       number = {1},
        pages = {295-307},
          doi = {10.1111/j.1365-2966.2008.13223.x},
archivePrefix = {arXiv},
       eprint = {0709.1544},
 primaryClass = {astro-ph},
       adsurl = {https://ui.adsabs.harvard.edu/abs/2008MNRAS.387..295A}
}

@ARTICLE{Jiang+12,
       author = {{Jiang}, Yan-Fei and {Stone}, James M. and {Davis}, Shane W.},
        title = "{A Godunov Method for Multidimensional Radiation Magnetohydrodynamics Based on a Variable Eddington Tensor}",
      journal = {\apjs},
         year = 2012,
        month = mar,
       volume = {199},
       number = {1},
          eid = {14},
        pages = {14},
          doi = {10.1088/0067-0049/199/1/14},
archivePrefix = {arXiv},
       eprint = {1201.2223},
 primaryClass = {astro-ph.HE},
       adsurl = {https://ui.adsabs.harvard.edu/abs/2012ApJS..199...14J}
}

@ARTICLE{HayesNorman03,
       author = {{Hayes}, John C. and {Norman}, Michael L.},
        title = "{Beyond Flux-limited Diffusion: Parallel Algorithms for Multidimensional Radiation Hydrodynamics}",
      journal = {\apjs},
         year = 2003,
        month = jul,
       volume = {147},
       number = {1},
        pages = {197-220},
          doi = {10.1086/374658},
archivePrefix = {arXiv},
       eprint = {astro-ph/0207260},
 primaryClass = {astro-ph},
       adsurl = {https://ui.adsabs.harvard.edu/abs/2003ApJS..147..197H}
}

@ARTICLE{Stone+92,
       author = {{Stone}, James M. and {Mihalas}, Dimitri and {Norman}, Michael L.},
        title = "{ZEUS-2D: A Radiation Magnetohydrodynamics Code for Astrophysical Flows in Two Space Dimensions. III. The Radiation Hydrodynamic Algorithms and Tests}",
      journal = {\apjs},
         year = 1992,
        month = jun,
       volume = {80},
        pages = {819},
          doi = {10.1086/191682},
       adsurl = {https://ui.adsabs.harvard.edu/abs/1992ApJS...80..819S}
}

@ARTICLE{Levermore84,
       author = {{Levermore}, C.~D.},
        title = "{Relating Eddington factors to flux limiters.}",
      journal = {\jqsrt},
         year = 1984,
        month = feb,
       volume = {31},
       number = {2},
        pages = {149-160},
          doi = {10.1016/0022-4073(84)90112-2},
       adsurl = {https://ui.adsabs.harvard.edu/abs/1984JQSRT..31..149L}
}

@ARTICLE{Sadowski+13,
       author = {{S{\k{a}}dowski}, Aleksander and {Narayan}, Ramesh and {Tchekhovskoy}, Alexander and {Zhu}, Yucong},
        title = "{Semi-implicit scheme for treating radiation under M1 closure in general relativistic conservative fluid dynamics codes}",
      journal = {\mnras},
         year = 2013,
        month = mar,
       volume = {429},
       number = {4},
        pages = {3533-3550},
          doi = {10.1093/mnras/sts632},
archivePrefix = {arXiv},
       eprint = {1212.5050},
 primaryClass = {astro-ph.HE},
       adsurl = {https://ui.adsabs.harvard.edu/abs/2013MNRAS.429.3533S}
}

@ARTICLE{Krumholz+07b,
       author = {{Krumholz}, Mark R. and {Klein}, Richard I. and {McKee}, Christopher F. and {Bolstad}, John},
        title = "{Equations and Algorithms for Mixed-frame Flux-limited Diffusion Radiation Hydrodynamics}",
      journal = {\apj},
         year = 2007,
        month = sep,
       volume = {667},
       number = {1},
        pages = {626-643},
          doi = {10.1086/520791},
archivePrefix = {arXiv},
       eprint = {astro-ph/0611003},
 primaryClass = {astro-ph},
       adsurl = {https://ui.adsabs.harvard.edu/abs/2007ApJ...667..626K}
}

@ARTICLE{Brutman+24,
       author = {{Brutman}, Yuval and {Steinberg}, Elad and {Balberg}, Shmuel},
        title = "{The Primary Flare Following a Stellar Collision in a Galactic Nucleus}",
      journal = {\apjl},
         year = 2024,
        month = oct,
       volume = {974},
       number = {1},
          eid = {L22},
        pages = {L22},
          doi = {10.3847/2041-8213/ad808f},
archivePrefix = {arXiv},
       eprint = {2408.16383},
 primaryClass = {astro-ph.HE},
       adsurl = {https://ui.adsabs.harvard.edu/abs/2024ApJ...974L..22B}
}

@ARTICLE{Huang+25,
       author = {{Huang}, Xiaoshan and {Linial}, Itai and {Jiang}, Yan-Fei},
        title = "{Multiband Emission from Star─Disk Collisions and Implications for Quasiperiodic Eruptions}",
      journal = {\apj},
         year = 2025,
        month = nov,
       volume = {993},
       number = {2},
          eid = {186},
        pages = {186},
          doi = {10.3847/1538-4357/ae07ca},
archivePrefix = {arXiv},
       eprint = {2506.11231},
 primaryClass = {astro-ph.HE},
       adsurl = {https://ui.adsabs.harvard.edu/abs/2025ApJ...993..186H}
}

@ARTICLE{Bonnerot+21,
       author = {{Bonnerot}, Cl{\'e}ment and {Lu}, Wenbin and {Hopkins}, Philip F.},
        title = "{First light from tidal disruption events}",
      journal = {\mnras},
         year = 2021,
        month = jul,
       volume = {504},
       number = {4},
        pages = {4885-4905},
          doi = {10.1093/mnras/stab398},
archivePrefix = {arXiv},
       eprint = {2012.12271},
 primaryClass = {astro-ph.HE},
       adsurl = {https://ui.adsabs.harvard.edu/abs/2021MNRAS.504.4885B}
}

@ARTICLE{Grudic+21,
       author = {{Grudi{\'c}}, Michael Y. and {Guszejnov}, D{\'a}vid and {Hopkins}, Philip F. and {Offner}, Stella S.~R. and {Faucher-Gigu{\`e}re}, Claude-Andr{\'e}},
        title = "{STARFORGE: Towards a comprehensive numerical model of star cluster formation and feedback}",
      journal = {\mnras},
         year = 2021,
        month = sep,
       volume = {506},
       number = {2},
        pages = {2199-2231},
          doi = {10.1093/mnras/stab1347},
archivePrefix = {arXiv},
       eprint = {2010.11254},
 primaryClass = {astro-ph.IM},
       adsurl = {https://ui.adsabs.harvard.edu/abs/2021MNRAS.506.2199G}
}

@ARTICLE{Krumholz+07a,
       author = {{Krumholz}, Mark R. and {Klein}, Richard I. and {McKee}, Christopher F.},
        title = "{Radiation-Hydrodynamic Simulations of Collapse and Fragmentation in Massive Protostellar Cores}",
      journal = {\apj},
         year = 2007,
        month = feb,
       volume = {656},
       number = {2},
        pages = {959-979},
          doi = {10.1086/510664},
archivePrefix = {arXiv},
       eprint = {astro-ph/0609798},
 primaryClass = {astro-ph},
       adsurl = {https://ui.adsabs.harvard.edu/abs/2007ApJ...656..959K}
}

@ARTICLE{Thompson+15,
       author = {{Thompson}, Todd A. and {Fabian}, Andrew C. and {Quataert}, Eliot and {Murray}, Norman},
        title = "{Dynamics of dusty radiation-pressure-driven shells and clouds: fast outflows from galaxies, star clusters, massive stars, and AGN}",
      journal = {\mnras},
         year = 2015,
        month = may,
       volume = {449},
       number = {1},
        pages = {147-161},
          doi = {10.1093/mnras/stv246},
archivePrefix = {arXiv},
       eprint = {1406.5206},
 primaryClass = {astro-ph.GA},
       adsurl = {https://ui.adsabs.harvard.edu/abs/2015MNRAS.449..147T}
}

@ARTICLE{YuanNarayan14,
       author = {{Yuan}, Feng and {Narayan}, Ramesh},
        title = "{Hot Accretion Flows Around Black Holes}",
      journal = {\araa},
         year = 2014,
        month = aug,
       volume = {52},
        pages = {529-588},
          doi = {10.1146/annurev-astro-082812-141003},
archivePrefix = {arXiv},
       eprint = {1401.0586},
 primaryClass = {astro-ph.HE},
       adsurl = {https://ui.adsabs.harvard.edu/abs/2014ARA&A..52..529Y}
}

@BOOK{KippenhahnWeigert13,
       author = {{Kippenhahn}, Rudolf and {Weigert}, Alfred and {Weiss}, Achim},
        title = "{Stellar Structure and Evolution}",
         year = 2013,
          doi = {10.1007/978-3-642-30304-3},
       adsurl = {https://ui.adsabs.harvard.edu/abs/2013sse..book.....K}
}

@ARTICLE{Jiang+19,
       author = {{Jiang}, Yan-Fei and {Stone}, James M. and {Davis}, Shane W.},
        title = "{Super-Eddington Accretion Disks around Supermassive Black Holes}",
      journal = {\apj},
         year = 2019,
        month = aug,
       volume = {880},
       number = {2},
          eid = {67},
        pages = {67},
          doi = {10.3847/1538-4357/ab29ff},
archivePrefix = {arXiv},
       eprint = {1709.02845},
 primaryClass = {astro-ph.HE},
       adsurl = {https://ui.adsabs.harvard.edu/abs/2019ApJ...880...67J}
}

@ARTICLE{Fabian99,
       author = {{Fabian}, A.~C.},
        title = "{The obscured growth of massive black holes}",
      journal = {\mnras},
         year = 1999,
        month = oct,
       volume = {308},
       number = {4},
        pages = {L39-L43},
          doi = {10.1046/j.1365-8711.1999.03017.x},
archivePrefix = {arXiv},
       eprint = {astro-ph/9908064},
 primaryClass = {astro-ph},
       adsurl = {https://ui.adsabs.harvard.edu/abs/1999MNRAS.308L..39F}
}

@article{BiCGSTAB,
author = {van der Vorst, H. A.},
title = {Bi-CGSTAB: A Fast and Smoothly Converging Variant of Bi-CG for the Solution of Nonsymmetric Linear Systems},
journal = {SIAM Journal on Scientific and Statistical Computing},
volume = {13},
number = {2},
pages = {631-644},
year = {1992},
doi = {10.1137/0913035},

URL = { 
    
        https://doi.org/10.1137/0913035
    
    

},
eprint = { 
    
        https://doi.org/10.1137/0913035
    
    

}
}

@article{DENSMORE20126924,
title = {A hybrid transport-diffusion Monte Carlo method for frequency-dependent radiative-transfer simulations},
journal = {Journal of Computational Physics},
volume = {231},
number = {20},
pages = {6924-6934},
year = {2012},
issn = {0021-9991},
doi = {https://doi.org/10.1016/j.jcp.2012.06.020},
url = {https://www.sciencedirect.com/science/article/pii/S002199911200335X},
author = {Jeffery D. Densmore and Kelly G. Thompson and Todd J. Urbatsch}
}

@Article{Lowrie2008,
author={Lowrie, Robert B.
and Edwards, Jarrod D.},
title={Radiative shock solutions with grey nonequilibrium diffusion},
journal={Shock Waves},
year={2008},
month={Jul},
day={01},
volume={18},
number={2},
pages={129-143},
issn={1432-2153},
doi={10.1007/s00193-008-0143-0},
url={https://doi.org/10.1007/s00193-008-0143-0}
}

@article{ferguson_nonrelativistic_2017,
	title = {Nonrelativistic grey {Sn}-transport radiative-shock solutions},
	volume = {23},
	issn = {1574-1818},
	url = {https://www.sciencedirect.com/science/article/pii/S1574181817300149},
	doi = {https://doi.org/10.1016/j.hedp.2017.02.010},
	journal = {High Energy Density Physics},
	author = {Ferguson, J. M. and Morel, J. E. and Lowrie, R. B.},
	year = {2017},
	pages = {95--114},
}

@article{malyali_transient_2024,
	title = {Transient fading {X}-ray emission detected during the optical rise of a tidal disruption event},
	volume = {531},
	issn = {0035-8711},
	url = {https://doi.org/10.1093/mnras/stae927},
	doi = {10.1093/mnras/stae927},
	number = {1},
	journal = {Monthly Notices of the Royal Astronomical Society},
	author = {Malyali, A and Rau, A and Bonnerot, C and Goodwin, A J and Liu, Z and Anderson, G E and Brink, J and Buckley, D A H and Merloni, A and Miller-Jones, J C A and Grotova, I and Kawka, A},
	month = apr,
	year = {2024},
	note = {\_eprint: https://academic.oup.com/mnras/article-pdf/531/1/1256/57801387/stae927.pdf},
	pages = {1256--1275},
}

@article{derei_non-equilibrium_2024,
	title = {The non-equilibrium {Marshak} wave problem in non-homogeneous media},
	volume = {36},
	issn = {1070-6631},
	url = {https://doi.org/10.1063/5.0244247},
	doi = {10.1063/5.0244247},
	number = {12},
	journal = {Physics of Fluids},
	author = {Derei, Nitay and Balberg, Shmuel and Heizler, Shay I. and Steinberg, Elad and McClarren, Ryan G. and Krief, Menahem},
	month = dec,
	year = {2024},
	note = {\_eprint: https://pubs.aip.org/aip/pof/article-pdf/doi/10.1063/5.0244247/20299761/127149\_1\_5.0244247.pdf},
	pages = {127149},
}

@article{krief_self-similar_2024,
	title = {Self-similar solutions for the non-equilibrium nonlinear supersonic {Marshak} wave problem},
	volume = {36},
	issn = {1070-6631},
	url = {https://doi.org/10.1063/5.0186666},
	doi = {10.1063/5.0186666},
	number = {1},
	journal = {Physics of Fluids},
	author = {Krief, Menahem and McClarren, Ryan G.},
	month = jan,
	year = {2024},
	note = {\_eprint: https://pubs.aip.org/aip/pof/article-pdf/doi/10.1063/5.0186666/18293668/017108\_1\_5.0186666.pdf},
	pages = {017108},
}

@article{steinberg_streamdisk_2024,
	title = {Stream–disk shocks as the origins of peak light in tidal disruption events},
	volume = {625},
	issn = {1476-4687},
	url = {https://doi.org/10.1038/s41586-023-06875-y},
	doi = {10.1038/s41586-023-06875-y},
	number = {7995},
	journal = {Nature},
	author = {Steinberg, Elad and Stone, Nicholas C.},
	month = jan,
	year = {2024},
	pages = {463--467},
}

@book{Castor_2004, place={Cambridge}, title={Radiation Hydrodynamics}, publisher={Cambridge University Press}, author={Castor, John I.}, year={2004}}

@article{yalinewich_rich_2015,
	title = {{RICH}: {OPEN}-{SOURCE} {HYDRODYNAMIC} {SIMULATION} {ON} {A} {MOVING} {VORONOI} {MESH}},
	volume = {216},
	url = {https://doi.org/10.1088/0067-0049/216/2/35},
	doi = {10.1088/0067-0049/216/2/35},
	number = {2},
	journal = {The Astrophysical Journal Supplement Series},
	author = {Yalinewich, Almog and Steinberg, Elad and Sari, Re’em},
	month = feb,
	year = {2015},
	note = {Publisher: The American Astronomical Society},
	pages = {35},
}

@book{mihalas_foundations_1999,
	series = {Dover {Books} on {Physics}},
	title = {Foundations of {Radiation} {Hydrodynamics}},
	isbn = {978-0-486-40925-2},
	url = {https://books.google.co.il/books?id=f75C_GN9KZwC},
	publisher = {Dover Publications},
	author = {Mihalas, D. and Weibel-Mihalas, B.},
	year = {1999},
	lccn = {99042987},
}

@ARTICLE{Pomeraning-Levermore,
       author = {{Levermore}, C.~D. and {Pomraning}, G.~C.},
        title = "{A flux-limited diffusion theory}",
      journal = {\apj},
         year = 1981,
        month = aug,
       volume = {248},
        pages = {321-334},
          doi = {10.1086/159157},
       adsurl = {https://ui.adsabs.harvard.edu/abs/1981ApJ...248..321L}
}

@article{Superbee,
	title = {Characteristic-{Based} {Schemes} for the {Euler} {Equations}},
	volume = {18},
	issn = {1545-4479},
	url = {https://www.annualreviews.org/content/journals/10.1146/annurev.fl.18.010186.002005},
	doi = {https://doi.org/10.1146/annurev.fl.18.010186.002005},
	number = {Volume 18, 1986},
	journal = {Annual Review of Fluid Mechanics},
	author = {Roe, P L},
	year = {1986},
	note = {Publisher: Annual Reviews
Type: Journal Article},
	pages = {337--365},
}

@article{fleck_implicit_1971,
	title = {An implicit {Monte} {Carlo} scheme for calculating time and frequency dependent nonlinear radiation transport},
	volume = {8},
	issn = {0021-9991},
	url = {https://www.sciencedirect.com/science/article/pii/0021999171900155},
	doi = {https://doi.org/10.1016/0021-9991(71)90015-5},
	number = {3},
	journal = {Journal of Computational Physics},
	author = {Fleck, J. A. and Cummings, J. D.},
	year = {1971},
	pages = {313--342},
}

@book{polyanin_handbook_2001,
	address = {London},
	title = {Handbook of {First}-{Order} {Partial} {Differential} {Equations}},
	isbn = {978-0-429-17296-0},
	publisher = {CRC Press},
	author = {Polyanin, Andrei D. and Zaitsev, Valentin F. and Moussiaux, Alain},
	month = nov,
	year = {2001},
	doi = {10.1201/b16828},
}

@article{CLARK1987311,
title = {Computing multigroup radiation integrals using polylogarithm-based methods},
journal = {Journal of Computational Physics},
volume = {70},
number = {2},
pages = {311-329},
year = {1987},
issn = {0021-9991},
doi = {https://doi.org/10.1016/0021-9991(87)90185-9},
url = {https://www.sciencedirect.com/science/article/pii/0021999187901859},
author = {Bradley A Clark}
}

@article{lewis2025fast,
  title={Fast, accurate numerical evaluation of incomplete Planck integrals},
  author={Lewis, Whit and McClarren, Ryan G},
  journal={Annals of Nuclear Energy},
  volume={218},
  pages={111374},
  year={2025},
  publisher={Elsevier}
}

@article{krief2018new,
  title={A new implementation of the STA method for the calculation of opacities of local thermodynamic equilibrium plasmas},
  author={Krief, Menahem and Feigel, Alexander and Gazit, Doron},
  journal={Atoms},
  volume={6},
  number={3},
  pages={35},
  year={2018},
  publisher={MDPI}
}

@article{krief2015variance,
  title={Variance and shift of transition arrays for electric and magnetic multipole transitions},
  author={Krief, Menahem and Feigel, Alexander},
  journal={High Energy Density Physics},
  volume={17},
  pages={254--262},
  year={2015},
  publisher={Elsevier}
}

@article{krief2015effect,
  title={The effect of first order superconfiguration energies on the opacity of hot dense matter},
  author={Krief, Menahem and Feigel, Alexander},
  journal={High Energy Density Physics},
  volume={15},
  pages={59--66},
  year={2015},
  publisher={Elsevier}
}

@article{krief2018effect,
  title={The effect of ionic correlations on radiative properties in the solar interior and terrestrial experiments},
  author={Krief, Menahem and Kurzweil, Yair and Feigel, Alexander and Gazit, Doron},
  journal={The Astrophysical Journal},
  volume={856},
  number={2},
  pages={135},
  year={2018},
  publisher={IOP Publishing}
}

@techreport{Till_T4,
  author       = {Till, Andrew Thomas},
  title        = {Discretization Writeup for Grey Flux-Limited Radiation Diffusion},
  institution  = {Los Alamos National Laboratory (LANL)},
  doi          = {10.2172/1716739},
  url          = {https://www.osti.gov/biblio/1716739},
  place        = {United States},
  year         = {2020},
  month        = {11}
}

@ARTICLE{2005ApJ...622..759G,
       author = {{G{\'o}rski}, K.~M. and {Hivon}, E. and {Banday}, A.~J. and {Wandelt}, B.~D. and {Hansen}, F.~K. and {Reinecke}, M. and {Bartelmann}, M.},
        title = "{HEALPix: A Framework for High-Resolution Discretization and Fast Analysis of Data Distributed on the Sphere}",
      journal = {\apj},
         year = 2005,
        month = apr,
       volume = {622},
       number = {2},
        pages = {759-771},
          doi = {10.1086/427976},
archivePrefix = {arXiv},
       eprint = {astro-ph/0409513},
 primaryClass = {astro-ph},
       adsurl = {https://ui.adsabs.harvard.edu/abs/2005ApJ...622..759G}
}

@article{Krumholz_2024,
    author = {He, Chong-Chong and Wibking, Benjamin D and Krumholz, Mark R},
    title = {A novel numerical method for mixed-frame multigroup radiation-hydrodynamics with GPU acceleration implemented in the quokka code},
    journal = {Monthly Notices of the Royal Astronomical Society},
    volume = {535},
    number = {4},
    pages = {3059-3076},
    year = {2024},
    month = {11},
    issn = {0035-8711},
    doi = {10.1093/mnras/stae2580},
    url = {https://doi.org/10.1093/mnras/stae2580},
    eprint = {https://academic.oup.com/mnras/article-pdf/535/4/3059/60817244/stae2580.pdf},
}

@article{steinberg2022multi,
  title={Multi-frequency implicit semi-analog Monte-Carlo (ISMC) radiative transfer solver in two-dimensions (without teleportation)},
  author={Steinberg, Elad and Heizler, Shay I},
  journal={Journal of Computational Physics},
  volume={450},
  pages={110806},
  year={2022},
  publisher={Elsevier}
}

@article{steinberg2023frequency,
  title={Frequency-Dependent Discrete Implicit Monte Carlo Scheme for the Radiative Transfer Equation},
  author={Steinberg, Elad and Heizler, Shay I},
  journal={Nuclear Science and Engineering},
  pages={1--13},
  year={2023},
  publisher={Taylor \& Francis}
}

@article{steinberg2022new,
  title={A new discrete implicit monte carlo scheme for simulating radiative transfer problems},
  author={Steinberg, Elad and Heizler, Shay I},
  journal={The Astrophysical Journal Supplement Series},
  volume={258},
  number={1},
  pages={14},
  year={2022},
  publisher={IOP Publishing}
}

@article{krief2024unified,
  title={A unified theory of the self-similar supersonic Marshak wave problem},
  author={Krief, Menahem and McClarren, Ryan G},
  journal={Physics of Fluids},
  volume={36},
  number={5},
  year={2024},
  publisher={AIP Publishing}
}

@article{wollaber2016four,
  title={Four decades of implicit Monte Carlo},
  author={Wollaber, Allan B},
  journal={Journal of Computational and Theoretical Transport},
  volume={45},
  number={1-2},
  pages={1--70},
  year={2016},
  publisher={Taylor \& Francis}
}

@article{gentile2001implicit,
  title={Implicit Monte Carlo diffusion—an acceleration method for Monte Carlo time-dependent radiative transfer simulations},
  author={Gentile, NA},
  journal={Journal of Computational Physics},
  volume={172},
  number={2},
  pages={543--571},
  year={2001},
  publisher={Elsevier}
}

@article{heizler2024accurate,
  title={Accurate reaction-diffusion limit to the spherical-symmetric Boltzmann equation},
  author={Heizler, Shay I and Krief, Menahem and Assaf, Michael},
  journal={Physical Review Research},
  volume={6},
  number={1},
  pages={L012023},
  year={2024},
  publisher={APS}
}

@article{giron2023solutions,
  title={Solutions of the converging and diverging shock problem in a medium with varying density},
  author={Giron, Itamar and Balberg, Shmuel and Krief, Menahem},
  journal={Physics of Fluids},
  volume={35},
  number={6},
  year={2023},
  publisher={AIP Publishing}
}

@article{krief2023piston,
  title={Piston driven shock waves in non-homogeneous planar media},
  author={Krief, Menahem},
  journal={Physics of Fluids},
  volume={35},
  number={4},
  year={2023},
  publisher={AIP Publishing}
}

@article{krief2021analytic,
  title={Analytic solutions of the nonlinear radiation diffusion equation with an instantaneous point source in non-homogeneous media},
  author={Krief, Menahem},
  journal={Physics of Fluids},
  volume={33},
  number={5},
  year={2021},
  publisher={AIP Publishing}
}

@article{giron2021solutions,
  title={Solutions of the imploding shock problem in a medium with varying density},
  author={Giron, Itamar and Balberg, Shmuel and Krief, Menahem},
  journal={Physics of Fluids},
  volume={33},
  number={6},
  year={2021},
  publisher={AIP Publishing}
}

@article{noebauer2019monte,
  title={Monte Carlo radiative transfer},
  author={Noebauer, Ulrich M and Sim, Stuart A},
  journal={Living Reviews in Computational Astrophysics},
  volume={5},
  number={1},
  pages={1},
  year={2019},
  publisher={Springer}
}

@techreport{Brunner2002Froms,
  title       = {Forms of Approximate Radiation Transport},
  author      = {Brunner, Thomas A.},
  institution = {Sandia National Laboratories},
  address     = {Albuquerque, NM and Livermore, CA},
  number      = {SAND2002-1778},
  year        = {2002},
  month       = {June},
  doi         = {10.2172/800993},
  url         = {https://www.osti.gov/biblio/800993}
}

@article{mcclarren2010theoretical,
  title={Theoretical aspects of the simplified Pn equations},
  author={McClarren, Ryan G},
  journal={Transport Theory and Statistical Physics},
  volume={39},
  number={2-4},
  pages={73--109},
  year={2010},
  publisher={Taylor \& Francis}
}

@article{till2018application,
  title={Application of linear multifrequency-grey acceleration to preconditioned Krylov iterations for thermal radiation transport},
  author={Till, Andrew T and Warsa, James S and Morel, Jim E},
  journal={Journal of Computational Physics},
  volume={372},
  pages={931--955},
  year={2018},
  publisher={Elsevier}
}

@article{winslow1995multifrequency,
  title={Multifrequency-gray method for radiation diffusion with Compton scattering},
  author={Winslow, Alan M},
  journal={Journal of Computational Physics},
  volume={117},
  number={2},
  pages={262--273},
  year={1995},
  publisher={Elsevier}
}
\bibliographystyle{aasjournal}

\appendix
\section{Computing Group Planck Integrals} \label{sec:computing_group_planck_integrals}
The normalized group Planck integrals $b_g$ were computed using a truncated polylogarithmic and Taylor series taken from Equations 32, 38 in \cite{CLARK1987311} , respectively. Taking the first 5 exponentials in the polylogarithmic series, the first 9 terms in the Taylor series and stitching them together at $x=2$ results in an accuracy of at least $10^{-5}$ (see Fig. 3 in \cite{CLARK1987311}). Concretely, we calculate $b_g$ using the following expression.
\begin{equation}
    b_g(T) = \Psi(\nu_{g+ 1/2}/k_BT,\ \nu_{g- 1/2}/k_BT)
\end{equation}

where  

\begin{equation}
\Psi\left(x_{1},x_{2}\right)=\begin{cases}
1+\frac{15}{\pi^{4}}\left(\Pi\left(x_{2}\right)-\Pi\left(x_{1}\right)\right), & x_{2}>2>x_{1}\\
\frac{15}{\pi^{4}}\left(\Pi\left(x_{2}\right)-\Pi\left(x_{1}\right)\right), & \text{else}
\end{cases}
\end{equation}

\begin{equation}
\Pi\left(x\right)=\begin{cases}
-x^{3}\sum_{l=1}^{5}\frac{e^{-lx}}{l}-3x^{2}\sum_{l=1}^{5}\frac{e^{-lx}}{l^{2}}-6x\sum_{l=1}^{5}\frac{e^{-lx}}{l^{3}}-6\sum_{l=1}^{5}\frac{e^{-lx}}{l^{4}}, & x>2\\
\frac{x^{3}}{3}-\frac{x^{4}}{8}+\frac{x^{5}}{60}-\frac{x^{7}}{5040}+\frac{x^{9}}{272160}, & \text{else}
\end{cases}
\end{equation}
 A detailed survey and analysis of the evaluation of incomplete Planck integrals was recently published by \cite{lewis2025fast}.
\end{document}